\documentclass{amsart}\usepackage{pb-diagram}

\input epsf.tex
\font \fraktur     = eufm10 scaled \magstephalf

\def\aci{a.c.i.\ system}
\def\BT{B\"acklund transformation}

\def\a{\alpha}
\def\va{\vec\a}
\def\b{\beta}

\def\l{\lambda}

\def\Px{{\mathcal P}}        
\def\Pg{\Px_{2g+1}}          
\def\Ug{{\mathcal C}_g}      
\def\Ugb{\bar{\mathcal C}_g} 
\def\F{{\mathcal F}}
\def\H{{\mathcal H}}         
\def\L{{\mathcal L}}         
\def\Lt{{\tilde L}}          
\def\D{{\mathcal D}}         
\def\Dt{{\tilde \D}}         
\def\LLt{{\tilde \L}}        
\def\O{{\mathcal O}}         

\font \fraktur=eufm10 scaled \magstephalf
\def \frak#1{\hbox{\fraktur #1}}
\def \sllie{{\frak s}{\frak l}}

\def\C{\mbox{$\mathbb C$}}
\def\Q{\mbox{$\mathbb Q$}}

\def\P{\mbox{$\mathbb P$}}

\def\Cur{{\Gamma}}    
\def\Cb{\bar\Cur}     

\def\Mh{{\hat M}}
\def\pt{{\tilde p}}
\def\qt{{\tilde q}}
\def\Pb{\{\cdot\,,\cdot\}}
\def\PB#1#2{\,\{#1\stackrel{\otimes}{,}#2\}\,}
\def\pp#1#2{\frac{\partial#1}{\partial#2}}
\def\si{\sum_{i=1}^g}
\def\sip{\sum_{i=0}^{g}}
\def\t{{\otimes}}
\def\utt{{\tilde {\tilde u}}}
\def\ut{{\tilde u}}
\def\vt{{\tilde v}}
\def\wt{{\tilde w}}
\def\Ut{{\tilde U}}
\def\Vt{{\tilde V}}
\def\Wt{{\tilde W}}
\def\xt{{\tilde x}}
\def\yt{{\tilde y}}
\def\zt{{\tilde z}}
\def\uvw{(u(x),\,v(x),\,w(x))}
\def\UVW{(U(x),\,V(x),\,W(x))}
\def\UVWt{(\tilde U(x),\,\tilde V(x),\,\tilde W(x))}
\def\uvwt{(\ut(x),\,\vt(x),\,\wt(x))}
\def\ddt{\frac{d}{dt}\vert_{t=0}}

\def\Ad{\mathop{\hbox{\rm Ad}}\nolimits}
\def\Id{\mathop{\hbox{\rm Id}}\nolimits}
\def\diag{\mathop{\hbox{\rm diag}}\nolimits}
\def\End{\mathop{\hbox{\rm End}}\nolimits}
\def\Jac{\mathop{\hbox{\rm Jac}}\nolimits}
\def\Pic{\mathop{\hbox{\rm Pic}}\nolimits}
\def\Span{\mathop{\hbox{\rm span}}\nolimits}
\def\Spec{\mathop{\hbox{\rm Spec}}\nolimits}
\def\Sym{\mathop{\hbox{\rm Sym}}\nolimits}

\def\mat#1#2#3#4{\left(\begin{array}{cc}#1&#2\\#3&#4\end{array}\right)}
\def\col#1#2{\left(\begin{array}{cc}#1\\#2\end{array}\right)}
\def\row#1#2{\left(\begin{array}{cc}#1&#2\end{array}\right)}

\begin{document}
\title[B\"acklund transformations]%
  {B\" acklund transformations for finite-dimensional
     integrable systems: a~geometric approach
}
\author{Vadim Kuznetsov}
  \address{Department of Applied Mathematics,
          University of Leeds,
          Leeds LS2 9JT, UK}
  \email{vadim@amsta.leeds.ac.uk}
\author{Pol Vanhaecke}
  \address{Universit\'e de Poitiers,
           D\'epartement de Math\'ematiques,
           T\'el\'eport 2,
           Boulevard Marie et Pierre Curie,
           BP 30179,
           F-86962 Futuroscope Chasseneuil Cedex}
  \email{Pol.Vanhaecke@mathlabo.univ-Poitiers.fr}
\keywords{Integrable Systems, B\"acklund transformations}
\subjclass{35Q58, 37J35, 58J72, 70H06}
%
\begin{abstract}
We present a geometric construction of \BT s and discretizations for a large class of
algebraic completely integrable systems. To be more precise, we construct families of
\BT s, which are naturally parametrized by the points on the spectral curve(s) of the
system. The key idea is that a point on the curve determines, through the Abel-Jacobi
map, a vector on its Jacobian which determines a translation on the corresponding
level set of the integrals (the generic level set of an algebraic completely
integrable systems has a group structure). Globalizing this construction we find
(possibly multi-valued, as is very common for \BT s) maps which preserve the
integrals of the system, they map solutions to solutions and they are symplectic maps
(or, more generally, Poisson maps). We show that these have the spectrality property,
a property of \BT s that was recently introduced. Moreover, we recover \BT s
and discretizations which have up to now been constructed by ad-hoc methods, and we
find \BT s and discretizations for other integrable systems. We also introduce
another approach, using pairs of normalizations of eigenvectors of Lax operators and
we explain how our two methods are related through the method of separation of
variables.
\end{abstract}
\maketitle
\tableofcontents

\section{Introduction} 

\label{intro}
The theory of integrable maps got a boost, if was not virtually
(re)started, a decade ago, when Veselov developed a theory of Lagrange
correspondences \cite{Ves1}, \cite{Ves2}. Roughly speaking,
\emph{integrable maps} (also called \emph{integrable Lagrange
correspondences}) are symplectic multi-valued mappings which have enough
integrals of motion, this definition being a proper analog of the classical
Liouville integrability. In the main examples, studied by him and later by
others, the integrable maps that are constructed are time-discretizations
of some classical Liouville integrable systems (such as the Neumann system,
the geodesic flow on an ellipsoid, the Euler-Manakov top, the Toda lattice,
Calogero-Moser systems and other integrable families), see, for instance,
\cite{HKR1}, \cite{HKR2}, \cite{KS}, \cite{KSS}, \cite{NRK}, \cite{BLSY},
\cite{MV} and \cite{BVe}. It follows that these symplectic maps associate
to a given solution of the integrable system a new solution, a property
reminiscent of \BT s for soliton equations; thus, one speaks in this
context often of a \emph{\BT\ } for the integrable system.

Recently \cite{KS} a new property of \emph{spectrality} of \BT s was
introduced. Namely, it was observed that when one searches for the simplest
\BT s of an integrable system, then one actually finds a one-dimensional
family $\{B_\l\mid\l\in\C\}$ of them and, most importantly, that the
variable $\mu$ which is essentially the conjugate\footnote{Since $B_\l$ is
symplectic it is given by a canonical transformation $F_\l$, which depends
on $\l$. The conjugate of $\l$ is given by $\partial F_\l/\partial\l$.} to
$\l$ is bound to $\l$ by the equation of an algebraic curve (dependent on
the integrals), which is precisely the curve that appears in the
linearization (integration) of the integrable system. The term
\emph{spectrality} stems from the fact that these curves arise most often
as spectral curves, e.g.\, when the vector fields of the integrable system
are given by Lax equations.

The purpose of this paper is to present
a systematic construction of \BT s for a large class of integrable systems which
includes most classical integrable systems and many new ones. Some of the flavors of
our methods and results are:
\begin{enumerate}
  \item Our \BT s $B_\l$ are given by  explicit formulas rather than implicit equations;
  \item We find big families of maps: one can let the parameter $\l$ vary from
    one level manifold of the integrals to another;
  \item They are symplectic (or Poisson) with respect to several
    compatible symplectic (or Poisson) structures;
  \item Although our maps are $n$-valued (two-valued in the examples),
    they lead to single-valued maps on any level manifold of the integrals;
  \item The resulting multi-point maps will discretize a family of
    flows of the integrable system (and not just  a particular one).
  \item The maps (and their iterates) are defined over an extension
  field $\Q(\sqrt{p})$ of $\Q$, where $p$ depends on the initial
  conditions (values of the integrals) only.
\end{enumerate}
These properties imply that our \BT s are very well suited as symplectic integrators
for the underlying integrable systems (see \cite{Mar}).

Our methods will be restricted to those integrable systems (defined over $\C$) which
have ``good'' algebraic geometric properties. These
systems,
baptized algebraic completely integrable systems (\aci s) by Adler and van Moerbeke
(see \cite{AvM1}) have algebraic integrals and Poisson structures, and the generic
common level set of the integrals is an affine part of a complex algebraic torus
(\emph{Abelian variety}) on which the flow of the integrable vector fields evolves
linearly. A \BT\ $B_\l$, as defined above, will leave each such level set
invariant. But it is well-know that Abelian varieties are rigid in the sense that a
holomorphic map between Abelian varieties is a group automorphism, followed by a
translation. The automorphism group of an Abelian variety being finite, $B_\l$
consists of a pure translation if it depends effectively on $\l$ and is the identity
map for some value of $\l$. If one wants to construct \BT s, one may therefore be
tempted to prescribe for every level set a $g$-dimensional vector ($g$ is the
dimension of the level set) but one is certainly doomed to fail when one wants to
write down explicitly in algebraic coordinates the map which results from a
translation over this family of vectors.

When the Abelian varieties that appear in the \aci\ are Jacobians then there is a
special family of translations, given by pairs of points on the underlying algebraic
curve (the Jacobian of an algebraic curve of genus $g$ is a $g$-dimensional Abelian
variety). Using the \emph{explicit} correspondence between the points of phase space
and the points on a Jacobian (represented either as divisors or line bundles on the
underlying curve) we write down the meromorphic function on the curve that realizes
the linear equivalence
\begin{equation}\label{intro1}
  \D+P\sim_l \tilde\D +Q,
\end{equation}
where $P$ and $Q$ are the two points on the curve and the divisors $\D$ and
$\tilde\D$ are the two divisors which correspond to a generic point on phase space
and its image under the \BT\ (this function is unique up to a constant factor). When
expressed in terms of the phase variables this provides us with the map that gives
the desired translation over the element $[P-Q]$ of the Jacobian. If one fixes one of
the points, say $Q$, one recovers a $1$-dimensional family of maps, indexed by a
point $P$ on the curve. Notice that we can vary the points from one Jacobian to the
other; however, there is an unavoidable monodromy problem, which makes that the
points $P$ and $Q$ may get interchanged (leading to precisely the opposite vector,
hence the inverse \BT), thus leading to a two-valued map.

For example, for the ($g$-dimensional) Mumford system (see \cite{Van1}), phase space
is the affine space of all matrices $L(x)=\mat{v(x)}{w(x)}{u(x)}{-v(x)}$ where $u,v$
and $w$ are polynomials in $x$ with $u$ and $w$ monic and
  $$\deg v(x)<\deg u(x)=\deg w(x)-1=g.
  $$
The family of maps that we construct are given by the similarity
transformation
\begin{equation}
  L(x)\mapsto M(x) L(x) M^{-1}(x)
\end{equation}
with
\begin{equation}
  M(x)=\mat{\b }{x-\l_f+\b^2}{1}{\b },
\end{equation}
where $\b=\frac{\mu_f-v(\l_f)}{u(\l_f)}$ and $(\l_f,\mu_f)$ is the
chosen point $P$ (dependent on $f$) on the spectral curve
$y^2=f(x)=-\det L(x)$ and $Q$ is the point at infinity of this
curve. It is easy to see that these maps satisfy properties 1, 2, 4
and 6 above.

By a direct computation we find, in each example, a large class of Poisson maps. In
the case of the Mumford system for example we show that when $P$ varies such that its
first coordinate depends on the Casimirs of the Poisson structure only, then we get a
Poisson map, thereby establishing property~3.

When the level manifolds of the \aci\ are not Jacobians then they are, in all known
examples where the integrals are known explicitly, covers of Jacobians, and we get
\BT s in an implicit form, i.e., we get Lagrangian correspondences as in Veselov's
original paper \cite{Ves1}. See Paragraph \ref{HH_par} for an example. The same
applies to g.\aci s (a.c.i.\ in the generalized sense, see \cite{AvM2}). When the
level manifolds are more general Abelian algebraic groups (a.c.i.\ in the sense of
Mumford) then they are extensions of Abelian varieties by one or more copies of
$\C^*$ and our technique again applies, see Paragraphs \ref{gen_par} and
\ref{gen_ev_par} for examples.

When we let $Q\to P$ then we find at the first order a vector field
which is constant on every level manifold because $Q$ and $P$ depend
on the integrals only, so their restrictions to these level manifolds
are linear combinations of the integrable vector fields. They need not
be globally Hamiltonian, but we will present in our examples
one-parameter families of points $(P,Q)$ which lead to precisely the
integrable vector fields of the \aci\ (property 5). In these cases the
\BT s should be considered as discretizations of the integrable
system. Since these \BT s commute, by construction, one may think of
these as defining a discrete analog of an \aci.

Below we will also present another, but related, technique to
construct the maps that represent translations on the level manifolds
(assumed to be affine parts of Jacobians) of the integrals. For this
it is assumed that phase space is given by Lax operators. We choose
two different normalizations of the eigenvectors of the Lax operator,
leading to two different separations of variables. This results in a
map which is identical to the one that we constructed before. The
reason is that the two different normalizations, which lead to
linearly equivalent divisors, are chosen such that each has a
different fixed point in the resulting divisor; if we call these
points $P$ and $Q$ then we recover precisely the above linear
equivalence (\ref{intro1}), and hence leads to the same \BT.
%
%
\section{The Mumford system} 
%
\label{Mumford}
%
  \subsection{Translations on hyperelliptic Jacobians} 
%
\label{back}
For a fixed integer $g\geq1$ the phase space $M_g$ of the ($g$-dimensional)
Mumford system (see \cite{Mum}) is the affine space $M_g$ of Lax matrices
$L(x)$ of the form
\begin{equation*}
  L(x)=\mat{v(x)}{w(x)}{u(x)}{-v(x)},
\end{equation*}
where $u(x),\,v(x)$ and $w(x)$ are polynomials, subject to the
following constraints: $u(x)$ and $w(x)$ are monic and their degrees
are respectively $g$ and $g+1$; the degree of $v(x)$ is at most
$g-1$. Writing
\begin{align*}
  u(x)&=x^g+u_1x^{g-1}+\ldots+u_g, \\
  v(x)&=v_1x^{g-1}+\ldots+v_g, \\
  w(x)&=x^{g+1}+w_0x^{g}+\ldots+w_g,
\end{align*}
we can take the coefficients of these three polynomials as coordinates
on $M_g$. In particular we will sometimes denote points of $M_g$ by
triples $\uvw$. Let us denote by $\Px_n$ the affine space of polynomials
$f\in\C[x]$ which are monic and have degree $n$. We will usually view $\Pg$
(or, in the next section, $\Px_{2g+2}$) as the space of hyperelliptic
curves with equation $y^2=f(x)$; when all roots of $f$ are distinct
then such a curve is smooth and its genus is $g$. We denote such an
affine curve by $\Cur_f$ and denote its smooth compactification, which
is a compact Riemann surface, by~$\Cb_f$. It is well-known that every
compact hyperelliptic Riemann surface of genus $g$ is obtained in this
way. The surjective map $\chi:M_g\to\Pg$ defined by
\begin{equation}\label{mmap}
  \chi(L(x))=-\det L(x)=u(x)w(x)+v^2(x)
\end{equation}
is the moment map of an algebraic completely integrable system (\aci). This means in
the first place that there is a Poisson structure\footnote{There are in fact in the
present case many (compatible) Poisson structures which make the Mumford system into
an \aci, see \cite{PV} and Paragraph \ref{odd_poisson_section} below.} on $M_g$ with
respect to which $\chi^*({\O}(\Pg))$ is involutive (commutative for the Poisson
bracket). Secondly, it means that the tangent space to a generic fiber $\chi^{-1}(f)$
of $\chi$ is spanned by the Hamiltonian vector fields associated to this involutive
algebra; by the first condition these vector fields commute. Third, a generic fiber
of $\chi$ is an affine part of a commutative algebraic group; in the present case,
when the roots of $f$ are distinct then $\chi^{-1}(f)$ is an affine part of a complex
algebraic torus, namely it is isomorphic to the Jacobian of $\Cb_f$, minus its theta
divisor. Finally, it means that the flow of the commuting Hamiltonian vector fields
on each complex torus lifts to a linear flow on its universal covering space~$\C^g$.

It is convenient for our constructions to introduce the universal
curve $\Ug$ of $\Pg$. Intuitively speaking, $\Ug$ is constructed out
of $\Pg$ by replacing every point of $\Pg$ by the curve which it
represents. Explicitly, $\Ug$ can be represented as the affine
variety
\begin{equation*}
  \left\{(x,y,f)\mid x,y\in\C,\,f\in \Pg\hbox{ and } y^2=f(x)\right\};
\end{equation*}
the natural projection $\Ug\to\Pg$ will be denoted by $\pi$. The
partial compactification of $\pi:\Ug\to\Pg$, which is the
quasi-projective variety obtained by compactifying the fibers of $\pi$,
will be denoted as $\Ugb$ and we use the same notation $\pi$ for the
extension of $\pi$ to $\Ugb$.

The first useful observation that we make is that any section $\xi$ of
$\pi:\Ug\to\Pg$ leads to a family of transformations of phase space, where each
transformation restricts to a translation on every Jacobian of the system. This
follows from the fact that there is a natural section $\xi_\infty$ of
$\pi:\Ugb\to\Pg$, which is given by $\xi_\infty(f)=(\infty_f,f)$, where $\infty_f$ is
the unique point needed to compactify $\Cur_f$ into $\Cb_f$. Indeed, if $\xi$ is a
section of $\pi:\Cur_g\to\Pg$ then we get a commutative diagram \par
\begin{equation*}
  \begin{diagram}
    \node{\Ugb}
    \node{M_g}\arrow{w,t}{\rho}
              \arrow{sw,r}{\chi}\\
    \node{\Pg}\arrow{n,l}{\xi}
  \end{diagram}
\end{equation*}
where $\rho$ is defined as $\rho=\xi\circ\chi$ and we get a map
$B_{\xi}:M_g\to M_g$ by
\begin{equation}\label{BT_gen}
  \L\mapsto \L\otimes[\rho(\L)-\rho_\infty(\L)],
\end{equation}
($\rho_\infty=\xi_\infty\circ\chi$). In this definition we use the fact that a
generic point $L(x)$ of $M_g$ (more precisely: each point of any fiber $\chi^{-1}(f)$
for which $\Cur_f$ is smooth) admits a natural interpretation as a holomorphic line
bundle $\L$ of degree $g$ over the Riemann surface $\Cb_f$, where $f=\chi(L(x))$;
thus $\L\in\Pic^g(\Cb_f)\cong\Jac(\Cur_f)$. Also, $[D]$ stands for the line bundle
associated to a divisor $D$. By construction, the restriction of $B_\xi$ to a generic
level $\chi^{-1}(f)$ of the moment map $\chi$ is a translation over
$[\xi(f)-\xi_\infty(f)]$. On the one hand this implies that $B_\xi$ is isospectral:
it leaves the fibers of $\chi$ invariant. On the other hand, translations in a
commutative group obviously preserve translation invariant vector fields, hence
$B_\xi$ leaves invariant all those vector fields on $M_g$ which restrict to
translation invariant vector fields on a generic fiber of $\chi$; in particular each
$B_\xi$ leaves the integrable vector fields of the Mumford system invariant. Notice
that it is unavoidable for such translation maps to have poles, because a non-zero
translation moves the theta divisor, hence every fiber of $\chi$ will have a divisor
of points which are sent out of phase space.

Our second observation is that the maps $B_\xi$ can be effectively computed. Indeed,
following Mumford (who attributes this construction to Jacobi) the above mentioned
interpretation of a generic element $L(x)\in M_g$ as a line bundle $\L$ can be
carried out explicitly as follows: to the point $L(x)=\uvw\in\chi^{-1}(f)$ we
associate the divisor $D=\sum_{i=1}^g(x_i,y_i)$ on $\Cur_f$ (hence the line bundle
$\L=[D]$ on $\Cb_f$, when $f$ is supposed to have no multiple roots) using the
following simple prescription:
\begin{align}
  &x_1,\dots,x_g \hbox{ are the zeros of $u(x)$,}
  \label{uvtodiv1}\\
  &y_i=v(x_i) \hbox{ for $i=1,\dots,g$}.
  \label{uvtodiv2}
\end{align}
Assuming $\uvw$ to be generic, we let $\Lt(x)=B_\xi(L(x))$ which we
also write as
\begin{equation*}
  \uvwt=B_\xi\uvw.
\end{equation*}
Since $\uvw$ is generic its image does indeed belong to $M_g$. We denote by $D$ the
divisor $\sum_{i=1}^g(x_i,y_i)$ given by (\ref{uvtodiv1}) and
(\ref{uvtodiv2}). According to (\ref{BT_gen}) the line bundle to which $[D]$ is
mapped is obtained by tensoring with $\left[\rho[D]-\rho_\infty[D]\right]$. We define
regular functions $\lambda$ and $\mu$ on $\Pg$ by $\xi(f)=(\lambda(f),\mu(f),f)$; in
order to simplify the notation we will write $\l_f$ and $\mu_f$ for $\l(f)$ and
$\mu(f)$.  Then (\ref{uvtodiv1}) and (\ref{uvtodiv2}) associate to $\uvwt$ the line
bundle $\tilde{\L}=[\tilde D]$ for which we have two different descriptions,
\begin{equation*}
  [\tilde D]= \left[\sum\nolimits_{i=1}^g(\xt_i,\yt_i)\right]
      = \left[\sum\nolimits_{i=1}^g(x_i,y_i)+(\l_f,\mu_f)-\infty_f\right].
\end{equation*}
The second equality expresses that $\sum\nolimits_{i=1}^g(\xt_i,\yt_i)+\infty_f$ and
$\sum\nolimits_{i=1}^g(x_i,y_i)+(\mu_f,\l_f)$ are linearly equivalent. This means
that there is a rational function (unique up to a non-zero constant) on $\Cb$ with
poles at $(x_i,y_i),\,(i=1,\ldots,g)$ and $(\l_f,\mu_f)$ and with a zero at
$\infty_f$. For any $\b\in\C$ we consider
\begin{equation}\label{F_odd}
  F(x,y)=\frac{y+v(x)+\b u(x)}{u(x)(x-\l_f)}.
\end{equation}
Taking a local parameter $t$ at $\infty_f$, such as $x=1/t^2$ and
$y=1/t^{2g+1}(1+O(t))$, we find that $F$ has a zero at $\infty_f$. Moreover, both the
numerator and denominator vanish at the points $(x_i,-y_i)$, hence it is sufficient
to have that $\b $ is such that the numerator vanishes at $(\l_f,-\mu_f)$ to have the
required function. Thus we take $\b $ to be given by
\begin{equation}\label{b_odd}
  \b =\frac{\mu_f-v(\l_f)}{u(\l_f)}=\frac{w(\l_f)}{\mu_f+v(\l_f)}.
\end{equation}
Notice that $\b$ depends on the phase variables; one may think of $\b$ itself as
being a phase variable, depending on the other phase variables (see also Paragraph
\ref{existence_parg} below). The zeros of $F$ on $\Cb_f$ are the points
$(\xt_i,\yt_i)$ and cannot be explicitly computed as such. However, the polynomials
$\uvwt$ to which they correspond, take a simple form. Consider
\begin{align*}
  (y-v(x)-\b u(x)) F(x,y) &= \frac{y^2-(v(x)+\b u(x))^2}{u(x)(x-\l_f)}\\
                       {}&= \frac{w(x)-2\b v(x)-\b^2u(x)}{x-\l_f}.
\end{align*}
Counting degrees we find that the last expression is monic of degree $g$ in $x$ and
is independent of $y$, hence it is $\prod_{i=1}^g(x-\xt_i)$, i.e., it is
$\ut(x)$. Thus we have obtained an explicit expression for the first component of
$B_\xi$:
\begin{equation}\label{ut_odd}
  \ut(x)=\frac{\b^2u(x)+2\b v(x)-w(x)}{\l_f-x}\,.
\end{equation}
We claim that the second component of $B_\xi$ is given by
\begin{align}
  \vt(x)&=-v(x)-\b  u(x)+\b  \ut(x)\cr
        &=\frac{\b(x-\l_f+\b^2)u(x)+(x-\l_f+2\b^2)v(x)-\b w(x)}{\l_f-x}\,.
  \label{vt_odd}
\end{align}
To show this, it suffices to verify that for generic $\uvw$ both sides take the same
value on $g$ different points (both sides are of degree at most $g-1$ in $x$). This
is easily done by using the points $(\xt_j,\yt_j)$ ($j=1,\dots,g$); just express that
$(\xt_j,\yt_j)\in\Cur_f$ and $F(\xt_j,\yt_j)=0$ for $1\leq j\leq g$, to find that
\begin{equation*}
  \tilde y_j=\vt(\tilde x_j)=-v(\tilde x_j) -\b u(\tilde x_j),
\end{equation*}
for $j=1,\ldots,g$. The formula for $\wt(x)$ follows from
\begin{equation*}
  \ut(x)\wt(x)+\vt^2(x)=f(x)=u(x)w(x)+v^2(x),
\end{equation*}
giving
\begin{equation}\label{wt_odd}
   \wt(x)=-\frac{(x-\l_f+\b^2)^2u(x)+2\b(x-\l_f+\b^2)v(x)-\b^2 w(x)}{\l_f-x}\,.
\end{equation}
Equations (\ref{ut_odd}), (\ref{vt_odd}) and (\ref{wt_odd}) give explicit formulas
for all maps $B_\xi$ ($\xi$ any section of $\Ug\to\Pg$). We will investigate the
poissonicity of the maps $B_\xi$ in Paragraph \ref{odd_poisson_section}.

We finish this section by rewriting $B_\xi$ in terms of matrices. Since $B_\xi$
preserves by construction the spectrum of the Lax matrix $L(x)$, it must be given by
a similarity transformation of $L(x)$,
\begin{equation}\label{Lt}
  \Lt(x)=M(x)L(x)M(x)^{-1}.
\end{equation}
It is easy to verify that such a matrix $M$ is given by the formula
\begin{equation}\label{M_odd}
  M(x)=\mat{\b }{x-\l_f+\b^2}{1}{\b }.
\end{equation}
Notice that $\det M(x) = \l_f-x$.

  \subsection{Poissonicity} %
%
\label{odd_poisson_section}
There are many (compabible) Poisson structures for the Mumford system on $M_g$ and
they can be obtained from a reduction of a natural class of $R$-brackets on the loop
algebra of $\sllie(2)$ (see \cite{PV}) or from (almost) canonical brackets on the
linearizing variables (see \cite{Van2}). Explicitly, there is a Poisson structure for
any univariate polynomial $\varphi(x)$ of degree at most $g$ and they are given by
the following Poisson brackets for the polynomials $u(x),v(x)$ and $w(x)$:
\begin{align}
  \{u(x),u(y)\}^\varphi&=\{v(x),v(y)\}^\varphi=0,\notag \\
  \{u(x),v(y)\}^\varphi&=\frac{u(x)\varphi(y)-u(y)\varphi(x)}{x-y},\notag\\
  \{u(x),w(y)\}^\varphi&=-2\frac{v(x)\varphi(y)-v(y)\varphi(x)}{x-y},\label{brackets}\\
  \{v(x),w(y)\}^\varphi&=\frac{w(x)\varphi(y)-w(y)\varphi(x)}{x-y}-u(x)\varphi(y),\notag\\
  \{w(x),w(y)\}^\varphi&=2\left(v(x)\varphi(y)-v(y)\varphi(x)\right).\notag
\end{align}
We will show that $B_\xi:\uvw\to\uvwt$ is a Poisson map for those sections $\xi$ for
which $\l$ depends on the Casimirs of $\Pb^\varphi$ only. More precisely, denoting
the algebra of Casimirs of $\Pb^\varphi$ by $Z^\varphi$ we assume in the sequel that
$\l$ factors over the canonical\footnote{$p$ is dual to the algebra homomorphism
$Z^\varphi\hookrightarrow \O(\Pg)$} map $p:\Pg\to\Spec Z^\varphi$, as in the
following diagram.
\begin{equation*}
  \begin{diagram}
    \node{}
    \node{\C}\\
    \node{\Spec Z^\varphi}
      \arrow{ne,l}{}
    \node{\Pg}
      \arrow{w,b}{p}
      \arrow{n,l}{\l}
  \end{diagram}
\end{equation*}
This assumption implies that $\l$ has trivial brackets with all phase variables;
notice that this does not imply that $\mu$ has trivial brackets with all phase
variables. One particular case of interest is when $\l$ is constant.

Using (\ref{brackets}) it can be shown by direct computation that the Poisson
brackets of the tilded variables are the same as those of the untilded variables ---
which proves that $B_\xi$ is a Poisson map --- but such computations are very long
and tedious. However, by using the Poisson bracket formalism that was introduced by
the Leningrad school these computations become feasible. In this
formalism one computes the $4\times4$ matrix $\PB{L(x)}{L(y)}$, which is defined
similarly as the tensor product of $L(x)$ and $L(y)$, but taking the Poisson bracket
of entries of $L(x)$ with entries of $L(y)$ instead of their product. Using this
notation (\ref{brackets}) can be written as
\begin{align}
\PB{L(x)}{L(y)}=&\left[r(x-y),L_1(x)\varphi(y)+\varphi(x)L_2(y)\right] \label{PB_len}\\
&-\left[\sigma\t\sigma,L_1(x)\varphi(y)-\varphi(x)L_2(y)\right]\notag
\end{align}
where $L_1(x)=L(x)\t\Id$, $L_2(y)=\Id\t L(y)$,
$\sigma=\mat0100
$ and
\begin{equation*}
  r(x)=-\frac1x\left(
    \begin{array}{cccc}
      1&0&0&0\\
      0&0&1&0\\
      0&1&0&0\\
      0&0&0&1\\
    \end{array}
  \right).
\end{equation*}
We need to verify that (\ref{PB_len}) also holds for the tilded variables, which
means, using $\Lt(x)=M(x)L(x)M(x)^{-1}$, that
\begin{align}
  \PB{M(x)&L(x)M(x)^{-1}}{M(y)L(y)M(y)^{-1}}=\notag\\
     &\left[r(x-y),M(x)L(x)M(x)^{-1}\t\Id\varphi(y)+\varphi(x)\Id\t
M(y)L(y)M(y)^{-1}\right]-
     \label{comp}\\
&\left[\sigma\t\sigma,M(x)L(x)M(x)^{-1}\varphi(y)\t\Id-\Id\t
\varphi(x)M(y)L(y)M(y)^{-1}\right]. \notag
\end{align}
In order to compute the left hand side of this equation we need explicit formulas for
$\PB{L(x)}{M(y)}$, for $\PB{M(x)}{L(y)}$ and for $\PB{M(x)}{M(y)}$. It is easy to see
that $\PB{M(x)}{M(y)}=0$. In order to find the other brackets we need the brackets of
$\beta$ with the other phase variables. They were computed from the definition
(\ref{b_odd}) of $\beta$, using the identity
$\{\mu_f^2-u(\l_f)w(\l_f)-v^2(\l_f),\cdot\}^\varphi=0$.
\begin{align*}
  \{u(x),\b\}^\varphi&=\frac{\mu_f\varphi(x)-\varphi(\l_f)(v(x)+\b
u(x))}{\mu_f(x-\l_f)},\\
\{v(x),\b\}^\varphi&=-\frac{2\mu_f\b\varphi(x)-\varphi(\l_f)(\b^2u(x)+w(x)-u(x)(x-\l_f))}
                              {2\mu_f(x-\l_f)},\\
  \{w(x),\b\}^\varphi&=-\frac{(\b^2+x-\l_f)\mu_f\varphi(x)+\varphi(\l_f)(\b^2v(x)-\b
                              w(x)-v(x)(x-\l_f))}{\mu_f(x-\l_f)}.
\end{align*}
Using these formulas it is easy to verify that
\begin{align*}
  \PB{L(x)}{M(y)}&=\left(\frac{\varphi(\l_f)}{2\mu_f}
  \left[L(x),M(x)^{-1}\frac{\partial
M}{\partial\b}\right]+\varphi(x)M(x)^{-1}\epsilon\right)\t
  \frac{\partial M}{\partial\b}\\
  \PB{M(x)}{L(y)}&=-\frac{\partial M}{\partial\b}\t\left(\frac{\varphi(\l_f)}
{2\mu_f}
  \left[L(y),M(y)^{-1}\frac{\partial M}{\partial\b}\right]
  +\varphi(y)M(y)^{-1}\epsilon\right),
\end{align*}
where $\epsilon=\diag(1,-1)$. For future use we note the following
identity
\begin{align}
  \Ad_{M(x)\t
  M(y)}(r(x-y)+\sigma\t\sigma)=r(x-y)+\sigma\t\sigma-\epsilon
  M(x)^{-1}\t \frac{\partial M}{\partial\b}M(y)^{-1}.\label{fund_id}
\end{align}
Since $\PB{M(x)}{M(y)}=0$ we get
\begin{align*}
  \PB{M(x)&L(x)M(x)^{-1}}{M(y)L(y)M(y)^{-1}}\\
    &=\Id\t M(y)\PB{M(x)}{L(y)} L(x)M(x)^{-1}\t M(y)^{-1}\\
    &\quad+M(x)\t\Id\PB{L(x)}{M(y)} M(x)^{-1}\t L(y)M(y)^{-1}\\
    &\quad+M(x)\t M(y)\PB{L(x)}{L(y)} M(x)^{-1}\t M(y)^{-1}\\
    &\quad-M(x)\t M(y)L(y)M(y)^{-1}\PB{L(x)}{M(y)} M(x)^{-1}\t M(y)^{-1}\\
    &\quad-M(x)L(x)M(x)^{-1}\t M(y)\PB{M(x)}{L(y)} M(x)^{-1}\t M(y)^{-1}.
\end{align*}
{}From here on the computation is straightforward: substitute the above expressions for
$\PB{L(x)}{L(y)}$, $\PB{L(x)}{M(y)}$ and $\PB{M(x)}{L(y)}$ and use, besides the
identity (\ref{fund_id}) the following formulas, valid for arbitrary matrices: $(A\t
B)(C\t D)=AC\t BD$ and $[A\t B,C\t D]=AC\t BD-CA\t DB$.  Notice that since each
expression is either linear in $\varphi(\l_f)$, in $\varphi(x)$ or in $\varphi(y)$
the computation can be split up in three shorter verifications.

It follows that $B_\xi:\uvw\to\uvwt$ is a Poisson map for those sections $\xi$ for
which $\l$ depends on the Casimirs of $\Pb^\varphi$ only. In view of the preceeding
section they are \BT s.

  \subsection{The existence of a section $\xi$} %
%
\label{existence_parg}

We have deliberately omitted the question of the {\it existence\/} of a (global)
section $\xi$ of $\pi:\Ug\to\Pg$. In fact it is easy to show that in the case of the
Mumford system such a (global) section does not exist. Indeed, let us suppose that
$\l:\Pg\to\C$ is given.  Since $\Pg$ consists of all monic polynomials of degree
$2g+1$ ($g\geq1$) the regular function $f\mapsto f(\l_f)$, defined on $\Pg$, is never
a constant map. Therefore it takes the value 0 at some point $f_0$, without being
identically zero on any neighborhood of $f_0$. If $\l$ is to be the first component
of a section $\xi$, i.e., $\xi(f)=(\l_f,\mu_f,f)$ then $\mu_f$ must be a regular map
on the affine space $\Pg$, satisfying $\mu_f^2=f(\l_f)$. On any neighborhood of $f_0$
this is however impossible. On the other hand it is clear that in a small
neighborhood $U$ of any $f\in\Pg$ a section $\xi$ exists: choose $\l:\Pg\to\C$ such
that $f(\l_{f})\neq0$. Thus the constructed \BT s should either be interpreted
semi-locally (i.e., on a neighborhood $\chi^{-1}(U)$ where $U$ is a neighborhood of a
fixed $f_0\in\Pg$), or one has to think of the \BT\ $B_{\xi}$ as a two-valued map. In
the latter interpretation it is worth to observe that the two translations which one
obtains are opposite to each other, as follows from
\begin{equation*}
  \left[(x,y)+(x,-y)-2\infty_f\right]=0,
\end{equation*}
valid for any $(x,y)\in\Cur_f$. On the one hand this implies that in a sense $B_\xi$
is its own inverse, on the other hand it implies that even an $n$-fold iteration of
$B_\xi$ is only 2-valued, not $2^n$-valued.

If one insists on having a \BT\ which is single-valued then one has to pass to a
cover of phase space, precisely as in the classical construction of Riemann surfaces
as the natural objects on which multi-valued algebraic functions become
single-valued. We wish to show now that this larger phase space inherits in fact a
Poisson structure and an \aci\ from the Mumford system, so that we have, in fact,
constructed a single-valued map for an \aci, which reduces to the Mumford system
after taking the quotient by an involution. Our arguments will be given here for the
Mumford system, but apply also to other systems, the involution being in general
replaced by a higher order automorphism. We fix a regular map $\l:\Pg\to\C$ and
define the following quasi projective variety,
\begin{equation*}
  M_g^\l=\left\{(u,v,w,\b)\mid
          (u,v,w)\in M_g,\,(\b u(\l_f)+v(\l_f))^2=f(\l_f),\,u(\l_f)\neq0\right\}.
\end{equation*}
The natural map $M_g^\l\to M_g$ is a two-fold ramified cover, and the
dynamics on this larger space, in particular the Poisson brackets of
$u,v$ and $w$ with $\b$ follow from the relation
\begin{equation*}
  \left\{(\b u(\l_f)+v(\l_f))^2-f(\l_f),\cdot\right\}=0,
\end{equation*}
(see \cite{Van1} for general constructions of this type). Since all our formulas for
the \BT\ were expressed regularly in terms of $u,v,w$ and $\b$ only, the \BT\ is
single-valued on this larger space. Obviously, the functions in involution of the
Mumford system lead to an algebra of functions in involution on the cover and, since
the dimension did not change, they still form an integrable system. To show that it
is actually an \aci\ we must investigate the nature of the generic fiber of the
moment map. For a generic $f\in\Pg$ we have that $f(\l_f)\neq0$. If we denote the two
square roots of $f(\l_f)$ by $\pm \mu_f$ then the fiber over $f$ is reducible and its
two components are given by
\begin{align*}
  &u(x)w(x)+v^2(x)=f(x),\\
  &\b u(\l_f)+v(\l_f)=\pm \mu_f.
\end{align*}
Notice that the two components do not intersect. Since we know that the variety in
$M_g$, given by $u(x)w(x)+v^2(x)=f(x)$ is an affine part of the Jacobian
$\Jac(\Cb_f)$, we find that each component is an affine part of $\Jac(\Cb_f)$; due to
the fact that $u(\l_f)=0$ along some divisor, the divisor which is removed in the
latter case is slightly larger than the one removed in the former case. Since the
lifted vector fields are also linear on these Jacobians this shows that the
integrable system that we have constructed is actually an \aci\ (with reducible
fibers).

Another way in which a global section $\xi$ in the case of the Mumford system can be
found is by passing to a subsystem, i.e., restricting phase space and its Poisson
structure to a hyperplane on which the algebra of functions in involution restricts
to an \aci. This smaller \aci\ is also universal for hyperelliptic curves in the
sense that, just as for the Mumford system, every hyperelliptic Jacobian (minus its
theta divisor) appears as one of the fibers of its moment map.
Suppose that $\F$ is an affine subspace of $\Pg$ and $\l$ is a regular
(or rational) function on $\F$ such that the $f(\l_f)=c$ where $c$ is
a constant, $c\in\C$. It can be shown that this implies that the map
$\l$ is constant. By adding $-c$ to all elements of $f$ we find that
all these polynomials have a common root $r$.  By replacing $x\to x+r$
in $f(x)$ this amounts to saying that up to isomorphism the only
reasonable subvariety of $M_g$ on which a global section $\xi$ can
exist is the subspace\footnote{This happens to be a Poisson subspace
for many (but not all) of the Poisson structures on $M_g$, see \cite{PV} or
Paragraph \ref{odd_poisson_section} above.} $M_g'$ of polynomials
$\uvw$ for which $u(0)w(0)+v^2(0)=0$; the map $\l$ must then be the
zero map, the section is given by $\l_f=(0,0,f)$ and the translation
on every fiber is given by $[(0,0)_f-\infty_f]$. Then $\b
=-v_g/u_g=w_g/v_g$ and the \BT\ takes the following form
\begin{align*}
  \ut_i&=w_{i-1}-2\frac{w_gv_{i-1}}{v_g}+\frac{w_gu_{i-1}}{u_g},\\
  \vt_i&=-v_i+\frac{v_g}{u_g}u_i-\frac{v_gw_{i-1}}{u_g}+
       2\frac{w_gv_{i-1}}{u_g}-\frac{v_gw_gu_{i-1}}{u_g^2}.
\end{align*}
Since $(0,0)_f$ is a Weierstrass point for any $f\in\F$ the divisor
$2((0,0)_f-\infty_f)$ is linearly equivalent to zero, in other words
$(0,0)_f-\infty_f$ is a half period (2-torsion point) on each
Jacobian. This explains why the two opposite translations are identical
and it shows that this \BT\ is an involution\footnote{The fact that
this \BT\ is an involution should not be confused with our earlier
claim that \emph{in a sense} the \BT\ is its own inverse.}.

  \subsection{Discretizations and continuum limits} %

\label{cont}
We now wish to show that the maps $B_\xi$ provide a discretization of
the Mumford system. Mumford constructs for every element of $\P^1$ a
vector field on $M_g$ which is translation invariant (linear) when
restricted to each fiber of $\chi$. His vector field corresponding to
$\infty$ is reconstructed here as the limit
\begin{equation*}
  \lim_{t\to0}\frac{B_{\xi_t}\uvw-\uvw}t
\end{equation*}
where $\xi_t:\Pg\to\Ug$ converges as $t\to0$ to the constant
section $\xi_\infty:\Pg\to\Ugb:f\mapsto\infty_f$. The limit taken here
is the one for which the sections $\xi_t(f)=(\l_f(t),\mu_f(t),f)$ take
the form
\begin{equation}\label{famsec}
  \xi_t(f)=\left(\frac1{t^2},\frac1{t^{2g+1}}\left(1+\frac{a_0}2t^2+
        O(t^4)\right),f\right);
\end{equation}
$a_0=u_1+w_0$ is the second coefficient of $f$, i.e.,
$f(x)=x^{2g+1}+a_0x^{2g}+\cdots$. Then
\begin{equation*}
  \b =\frac1t\left(1+\frac{w_0-u_1}2t^2+O(t^3)\right),
\end{equation*}
hence (\ref{ut_odd}), (\ref{vt_odd}) and (\ref{wt_odd}) take the form
\begin{align}
  \ut(x)&=u(x)+2tv(x)+O(t^2),\notag\\
  \vt(x)&=v(x)-t(w(x)-(x-u_1+w_0)u(x))+O(t^2),\label{example}\\
  \wt(x)&=w(x)-2t(x-u_1+w_0)v(x)+O(t^2).\notag
\end{align}
The coefficient of $t$ in (\ref{example}) is (up to a factor of 2)
precisely Mumford's vector field $X_\infty$ (see \cite{Mum} page 3.43).

Let us now turn to Mumford's general vector fields $X_a$
($a\in\P^1$). These vector fields have the property of being tangent
to the curves $P\mapsto[P+(g-1)\infty]$ at the points $(a,\pm b_f)$ on
every curve $f$ (here $b^2_f=f(a)$), which suggests that these more
general vector fields may be constructed by taking an appropriate
limit $(\l_f,\mu_f)\to(a,b_f)$ of the composition of two \BT s
corresponding to a shift
\begin{equation*}
  [(\l_f,\mu_f)-(a,b_f)]=[(\l_f,\mu_f)+(a,-b_f)-2\infty_f]
\end{equation*}
on each Jacobian. Our vector fields will be more general than
Mumford's vector fields because we allow $a_f$ to depend on $f$.
Concretely, we will first shift over $[(a_f,-b_f)-\infty_f]$ and then
over $[(\l_f(t),\mu_f(t))-\infty_f]$; the matrices going with these
transformations (as in (\ref{M_odd})) will be denoted by $P(x)$ and
$Q_t(x)$. Then
\begin{equation*}
  P(x)=\mat{-\b }{x-a_f+\b^2}{1}{-\b }
\end{equation*}
with
\begin{equation}\label{bf_odd}
  \b =\frac{b_f+v(a_f)}{u(a_f)}=\frac{w(a_f)}{b_f-v(a_f)};
\end{equation}
the transformed $L$ is denoted by $\Lt$ as in (\ref{Lt}). In
particular,
\begin{align}
  \ut(x)&=\frac{w(x)+2\b v(x)-\b^2u(x)}{x-a_f},\label{ut1}\\
  \vt(x)&=-v(x)+\b u(x)-\b \ut(x).\notag
\end{align}
Also,
\begin{equation*}
  Q_t(x)=\mat{\b (t)}{x-\l_f(t)+\b^2(t)}{1}{\b (t)}
\end{equation*}
with
\begin{equation*}
  \b (t)=\frac{\mu_f(t)-\vt(\l_f(t))}{\ut(\l_f(t))}.
\end{equation*}
Notice that $\b (0)=\b $ since $(\l_f(0),\mu_f(0))=(a_f,b_f)$.  Let
$M_t(x)=Q_t(x)P(x)$ be the matrix defining their composition. To the
deformation family $\Lt_t(x)=M_t(x)L(x)M_t^{-1}(x)$ there corresponds a
vector field on $M_g$, defined by
\begin{equation*}
  \frac{dL}{dt_{a_f}}(x)=\ddt\left(M_t(x)L(x)M_t^{-1}(x)\right).
\end{equation*}
In terms of $Q(x)$ this vector field is given by (a prime denotes a
derivative with respect to $t$)
\begin{equation*}
  \frac{dL}{dt_{a_f}}(x)=\left[M'_0(x)\,M_0(x)^{-1},L(x)\right]
                 =\left[Q'_0(x)\, Q_0^{-1}(x),L(x)\right].
\end{equation*}
We consider the family of sections $\xi_t=(\l_f(t),\mu_f(t),f)$ where
$\l_f(t)=a_f+t$ and $\mu_f(t)=\sqrt{f(a_f+t)}$. We will show below
that
\begin{equation}\label{calc}
  \b'(0)=\frac{u(a_f)}{2b_f}.
\end{equation}
Then
\begin{align*}
  Q_0'(x)Q_0^{-1}(x)&=-\frac1{2b_f(x-a_f)}\mat{u(a_f)}{2v(a_f)}0{u(a_f)}
           \mat{\b }{a_f-x-\b^2}{-1}{\b }\\
        &=\frac1{2b_f(x-a_f)}\mat{v(a_f)-b_f}{w(a_f)+u(a_f)(x-a_f)}{u(a_f)}
{-v(a_f)-b_f}.
\end{align*}
Removing a diagonal matrix from this matrix we get the following Lax
equations
\begin{equation*}
  \frac{dL}{dt_{a_f}}(x)=\frac1{2b_f}\left[\frac{L(a_f)}{x-a_f}+
         \mat0{u(a_f)}00,L(x)\right],
\end{equation*}
which reduces, when $a_f=a$ is chosen independently of $f$, to Mumford's vector
field $X_a$ (up to a factor $2b_f$ which can be absorbed in $t$).

Formula (\ref{calc}) remains to be shown.
\begin{align*}
  \b'(0)&=\ddt\frac{\mu_f(t)-\vt(\l_f(t))}{\ut(\l_f(t))}\\
          &=\ddt\frac{\mu_f(t)+v(a_f+t)-\b u(a_f+t)}
                      {w(a_f+t)+2\b v(a_f+t)-\b^2u(a_f+t)}t\\
          &=\lim_{t\to0}\frac{\mu_f(t)+v(a_f+t)-\b u(a_f+t)}
                      {w(a_f+t)+2\b v(a_f+t)-\b^2u(a_f+t)}\\
          &=\frac{\mu_f'(0)+v'(a_f)-\b u'(a_f)}
                      {w'(a_f)+2\b v'(a_f)-\b^2u'(a_f)}.
\end{align*}
Taking the derivative of $\mu^2_f(t)=u(\l_f(t))w(\l_f(t))+v^2(\l_f(t))$
at $t=0$ we obtain
\begin{equation*}
  \mu_f'(0)=\frac1{2b_f}\left(u(a_f)w'(a_f)+u'(a_f)w(a_f)+2v(a_f)v'(a_f)\right)
\end{equation*}
and $w(a_f)$ is easily eliminated from this equation by using
$w(a_f)=-2\b v(a_f)+\b^2u(a_f)$, a consequence of (\ref{ut1}). The
announced formula for $\b'(0)$ follows after substituting this value
of $\mu_f'(0)$, upon using (\ref{bf_odd}).

  \subsection{Normalizations of eigenvectors of Lax operators} 

In this section we describe another approach to \BT s and we explain how
the two approaches are related. For this approach we assume that the
\aci\  is given in Lax form.

Let us recall (see e.g., \cite{Gr}) that a generic Lax matrix
$L(x)\in\End(\C^{n+1})[x]$ defines a line bundle on the associated spectral
curve $\Cur:\det(L(x)-y\Id)=0$; generic means here that the affine
curve $\Cur$ is assumed smooth and that for generic $(x,y)\in\Cur$ the
eigenspace of $L(x)$ corresponding to the eigenvalue $y$ is
1-dimensional (both conditions are verified for the generic $L(x)$ of
the Mumford system). Assuming $L(x)$ to be generic we denote, as
before, by $\Cb$ the compact Riemann surface corresponding to
$\Cur$ and we consider the eigenvector map $\kappa:\Cb\to\P^{n}$, which
is defined, on the affine piece $\Cur$, by
\begin{equation*}
  L(x)\kappa(x,y)=y\kappa(x,y).
\end{equation*}
An explicit description of $\kappa$ on an affine piece of $\Cb$ is
given by the map
\begin{equation}\label{locrep}
  \kappa_i:(x,y)\mapsto (L(x)-y\Id)^\wedge_i
\end{equation}
where $1\leq i\leq n+1$ is arbitrary, $A^\wedge$ stands for the
adjoint of the matrix $A$ and $A_i$ stands for the $i$-th column of
$A$. More precisely, every $\kappa_i$ is defined on $\Cur\setminus
S_i$, where $S_i$ is a collection of points and
$\cap_iS_i=\emptyset$. We will see shortly that we need {\sl all\/}
local representatives $\kappa_i$ ($i=1,\dots, n+1$) of $\kappa$ for
our computations.  The line bundle $\L$, defined by $L(x)$, is given
by $\L=\kappa^*\H$, where $\H$ is the hyperplane bundle on $\P^{n}$.
The degree $d$ of $\L$ follows from
\begin{equation}\label{degreeL}
  \deg\L=\deg\kappa(\Cb)\deg\kappa.
\end{equation}
It is a basic fact that pulling back a section $s$ of $\H$ gives a
section $\kappa^* s$ whose zero locus is a divisor $D$ on $\Cur$ such
that $[D]=\kappa^* \H$ (see \cite{GH} Ch.\ 1.1). Since a section of $\H$ is
just a hyperplane, this gives us an explicit way to compute the line
bundle $\L\in\Pic^d(\Cb)$ from the Lax matrix:
\begin{equation}\label{Lbdef}
  \L=\left[\kappa^*(H\cap\kappa(\Cb))\right],
\end{equation}
where $H$ is any hyperplane in $\P^n$. Moreover, the
isomorphism $\Pic^d(\Cb)\cong\Jac(\Cb)$ is {\sl not\/} canonical and
depends on the choice of an element in $\Pic^{d-g}(\Cb)$, a fact that
we will now exploit to construct \BT s.

To do this we assume that the given $L(x)$ is generic in the above
sense; without loss of generality we may also assume that the image
curve $\kappa(\Cb)$ is non-degenerate (i.e., it is not contained in a
hyperplane). Our main assumption, which will be relaxed in Section
\ref{complex}, is that $\deg\L=g+n$. Since the hyperplane bundle $\H$
on $\P^n$ is the line bundle which corresponds to any hyperplane of
$\P^n$, fixing a section of $\H$ is equivalent to fixing a hyperplane
$H$ of $\P^n$. By non-degeneracy this can be done by fixing $n$ points
$p_i$ on $\Cb$ which are in general position, and asking that $H$ be
such that $\sum p_i\leq \kappa^*H$ (when all $p_i$ are different this
means that $H=\Span{\{}\kappa(p_i){\}}$). Let us take another
collection of $n$ points $\tilde p_i$ in general position. We denote
the corresponding hyperplane by $\tilde H$. If $\Lt(x)$ is another Lax
matrix, isospectral to $L(x)$, with corresponding map
$\tilde\kappa:\Cb\to\P^n$ then we will say that $\Lt(x)=B(L(x))$ if
\begin{equation}\label{BA_eq}
  \tilde\kappa^*(\tilde H\cap\tilde\kappa(\Cb))-\sum_{i=1}^n\tilde p_i=
  \kappa^*(H\cap\kappa(\Cb))-\sum_{i=1}^np_i.
\end{equation}
Notice that (\ref{BA_eq}) implies that
\begin{equation}\label{BT_alt}
  \LLt=\L\otimes
     [\tilde p_1-p_1]\otimes\cdots\otimes[\tilde p_n-p_n],
\end{equation}
where $\L$ is given by (\ref{Lbdef}) and $\LLt$ is defined
analogously. One notices that this equation is the $n$-point analog of
equation (\ref{BT_gen}). In fact, let us specialize this to the case
$n=1$ and globalize the construction to the phase space of the Mumford
system and recover exactly the \BT s that we have constructed before.

If $L(x)$ is a generic matrix of $M_g$ (the phase space of the Mumford
system) then $n=1$ and the two local representatives (\ref{locrep}) of
the eigenvalue map $\kappa$ are given by
\begin{equation*}
  \kappa_1:(x,y)\mapsto\col{-v(x)-y}{-u(x)} \qquad\hbox{and}\qquad
  \kappa_2:(x,y)\mapsto\col{-w(x)}{v(x)-y}.
\end{equation*}
A hyperplane $H$ of $\P$ is just a point: writing $\va=(r\colon s)$ we
find the following equations for the divisor
$D=\kappa^*(H\cap\kappa(\Cur_f))$:
\begin{align*}
  0&=(v(x)+y)r+u(x)s,\\
  0&=-w(x)r+(v(x)-y)s.
\end{align*}
The degree of the image curve being 1 it suffices to determine the degree of $\kappa$
to know the degree of the line bundle. Taking a $(r\colon s)$ generic, we easily find
precisely $g+1$ solutions hence $\deg\L=g+1$, showing that our main assumption is
satisfied for the Mumford system. Since $n=1$ we need to pick one point on every
curve $\Cb_f$ to represent $\L$ as an element of the Jacobian
$\Jac(\Cb_f)=\Pic^g(\Cb_f)$ and we need two points on every curve to construct a \BT\
as in (\ref{BA_eq}). We do this by picking the sections $\xi_\infty$ and $\xi$ which
were constructed in Paragraph \ref{back}. For the first choice, which corresponds to
picking the point $\infty_f$ at every curve, we find $\va_0=(0\colon1)$; we let this
choice correspond to the untilded variables. We let the second choice, which is given
by $\xi(f)=(\l_f,\mu_f,f)$, correspond to the tilded variables and we
find\footnote{Given $L(x)$ there are $g$ (resp.\ $g+1$) values $(\l,\mu)$ where the
first (resp.\ second) representation breaks down, i.e., it may be of the form
$\va=(0:0)$. For generic $L(x)$ those two sets of values are disjoint, in the
non-generic case it suffices to take a limit.}
\begin{equation*}
  \va=(\ut(\l_f):-\vt(\l_f)-\mu_f)=(\vt(\l_f)-\mu_f:\wt(\l_f)).
\end{equation*}
In order to simplify the computation we will write $\va$ as $(1:-\b )$; it will
follow later that this definition of $\b $ agrees with the one given in
(\ref{b_odd}). (\ref{BT_alt}) now expresses that the solutions of
\begin{equation*}
  u(x)=0,\qquad\qquad v(x)=y,
\end{equation*}
are the same as the solutions of
\begin{equation}\label{tilde_eq}
  \row{1}{-\b }\mat{-\vt(x)-y}{-\wt(x)}{-\ut(x)}{\vt(x)-y}=0,
\end{equation}
except that (\ref{tilde_eq}) also has $(\l_f,\mu_f)$ as a solution. If
we eliminate $y$ from (\ref{tilde_eq}) we find that
$\wt(x)+2\b \vt(x)-\b^2\ut(x)=0$ has as solutions $\l_f$ and the
roots of $u$, so
\begin{equation}\label{1stpart}
  u(x)=\frac{\b^2\ut(x)-2\b \vt(x)-\wt(x)}{\l_f-x}.
\end{equation}
In order to obtain the formula for $v(x)$ we take the first equation
in (\ref{tilde_eq}), $-\vt(x)-y+\b \ut(x)=0$ which has among its roots
the solutions of $u(x)=0,\ v(x)=y$. It follows that the same is true
for the polynomial $-\vt(x)-v(x)-\b u(x)+\b \ut(x)=0$, but since
this polynomial has degree less than $g$ it is zero, giving
\begin{equation}\label{2ndpart}
  v(x)=-\vt(x)-\b u(x)+\b \ut(x).
\end{equation}
If we express that $(\l_f,\mu_f)$ is a solution to (\ref{tilde_eq}),
then (\ref{2ndpart}) implies
\begin{equation*}
  \b=\frac{\vt(\l_f)+\mu_f}{\ut(\l_f)}=\frac{\mu_f-v(\l_f)}{u(\l_f)},
\end{equation*}
as in (\ref{b_odd}). It follows that formulas (\ref{1stpart}) and
(\ref{2ndpart}) describe exactly the maps $B_\xi$, given by
(\ref{ut_odd}) and (\ref{vt_odd}), in their inverse form.
Notice that we would have obtained an expression for the maps $B_\xi$
in their direct form by expressing that the solutions to
\begin{equation*}
  \ut(x)=0,\qquad\qquad \vt(x)=y,
\end{equation*}
are the same as the solutions of
\begin{equation}\label{tilde_eq_3}
  \row{1}{-\b }\mat{-v(x)-y}{-w(x)}{-u(x)}{v(x)-y}=0,
\end{equation}
except that (\ref{tilde_eq_3}) also has $(\l_f,-\mu_f)$ as a solution
(this follows from the linear equivalence
$(\l_f,\mu_f)+(\l_f,-\mu_f)\sim_l 2\infty_f$).

It follows from \cite{Mum} that the roots of the polynomial $u(x)$ lead to a
separation of variables. This is one separation of variables; another one
is given by the equations (\ref{tilde_eq}) for the
tilde-variables. Relating them by assuming that they have the same divisor
$D$ as a solution, we create a \BT \ which corresponds to a shift on each
Jacobian parametrized by a point $(\l_f,\mu_f)$ on its underlying curve
$\Cur_f$.  Thus, in the Lax approach, our construction of \BT s leads to
alternative separation of variables (given one separation of variables) and
given a pair of separations of variables we recover a \BT\ for the system.

  \subsection{Spectrality} %

We now come to a remarkable property of our \BT s, which was baptized
\emph{spectrality} by \cite{KS}.  In order to establish this property we will first
consider an isomorphism to another integrable system in which the Poisson structure
takes a simple form. We fix an irreducible monic polynomial $\varphi(x)$ of degree
$g$,
\begin{equation*}
  \varphi(x)=(x-a_1)(x-a_2)\cdots(x-a_g),
\end{equation*}
and we define an affine map $M_g\to\C^{3g+1}$ by
\begin{equation*}
  \frac{1}{\varphi(x)}(u(x),v(x),w(x))=\left(1+\si\frac{f_i}{x-a_i},\si\frac{h_i}
{x-a_i},
  x+e_0+\si\frac{e_i}{x-a_i}\right).
\end{equation*}
Explicitly, the map can be computed in terms of the coordinates
$e_0,\dots,h_g$ on $\C^{3g+1}$ by
\begin{equation*}
  f_i=\frac{u(a_i)}{\prod_{k\neq i} (a_i-a_k)}\,,\quad
  h_i=\frac{v(a_i)}{\prod_{k\neq i} (a_i-a_k)}\,,\quad
  e_i=\frac{w(a_i)}{\prod_{k\neq i} (a_i-a_k)}\,,
\end{equation*}
and $e_0=w_0-\si a_i$. Dividing both sides of the equations
(\ref{brackets}) by $\varphi(x)\varphi(y)$ and taking residues at $x=a_i$
and $y=a_j$ we find that the variables $\{h_i,e_i,f_i\}_{i=1}^g$ are
generators for the direct sum of $g$ copies of the Lie-Poisson algebra
of $\sllie(2)$: for $i,j=1,\ldots,g$, we have $\{h_i,h_j\}= \{f_i,f_j\}=
\{e_i,e_j\}=0$ and
\begin{equation}\label{sl2brackets}
  \{e_i,h_j\}=e_i\delta_{ij},\qquad
  \{h_i,f_j\}=f_i\delta_{ij},\qquad
  \{f_i,e_j\}=2h_i\delta_{ij}.
\end{equation}
Let us denote the Casimir element coming from the
$i$-th copy of $\sllie(2)$ by $C_i$, $C_i=h_i^2+e_if_i$. Then the equation
of the spectral curve looks as follows:
\begin{equation}\label{x-curve}
  \frac{y^2}{\varphi^2(x)}=\frac{f(x)}{\varphi^2(x)}=
  x+C_0+\sum_{i=1}^g\left(\frac{C_i}{(x-a_i)^2}+\frac{H_i}{x-a_i}\right),
\end{equation}
where
\begin{equation*}
  H_i=\sum_{j\neq i}\frac{2h_ih_j+e_if_j+e_jf_i}{a_i-a_j}+e_i+(a_i+e_0)f_i
\end{equation*}
and $C_0$ is an extra Casimir. If we define $\hat\mu_f=\mu_f/\varphi(\l_f)$ then
\begin{equation*}
  \hat\mu_f^2=\l_f+C_0+\sum_{i=1}^g\left(\frac{C_i}{(\l_f-a_i)^2}+
  \frac{H_i}{\l_f-a_i}\right),
\end{equation*}
and the relation (\ref{b_odd}) takes the form
\begin{equation}\label{3}
  \b=\frac{\hat\mu_f-\sum_i\frac{h_i}{\l_f-a_i}}{1+\sum_i\frac{f_i}{\l_f-a_i}}\,.
\end{equation}
Notice that on $\C^{3g+1}$ the Poisson structure is independent of
$\varphi$, but that the Hamiltonians are now dependent on the constants
$a_i$ which encode the Poisson structure $\Pb^\varphi$ on $M_g$. In fact,
the integrable system that we have obtained on $\C^{3g+1}$ is the first
member of the deformed Gaudin magnet hierarchy from \cite{EEKT} and
our \BT s for the Mumford system are easily rewritten as \BT s for this
system. Explicitly we find
\begin{align}
  \tilde f_i&=\frac{\b^2f_i+2\b h_i-e_i}{\l_f-a_i}\,,\notag\\
  \tilde h_i&=\frac{\b(a_i-\l_f+\b^2)f_i+(a_i-\l_f+2\b^2)h_i-\b
  e_i}{\l_f-a_i}\,,\label{i}\\
  \tilde e_i&=-\frac{(a_i-\l_f+\b^2)^2f_i+2\b(a_i-\l_f+\b^2)h_i-\b^2
  e_i}{\l_f-a_i}\,,\notag
\end{align}
where $\beta$ is given by (\ref{3}).

We fix a section $\xi$ of $\Ug\to\Pg$ and we assume, as before, that
$\l_f$ depends on the Casimirs of $\Pb^\varphi$ only, where
$\xi(f)=(\l_f,\mu_f)$. We restrict our \BT\ $B_\xi$ to a symplectic leaf of the Poisson
structure by
fixing generic values of all Casimirs $C_j$, $j=0,\ldots,g$. Then we have
only $2g$ independent (Darboux-type) variables, which we choose to be
$\{h_i,f_i\}_{i=1}^g$, we can express the $e_i$ variables in terms
of those (the expression for $e_0$ was computed from (\ref{x-curve})),
\begin{equation*}
  e_i=\frac{C_i-h_i^2}{f_i}\,,\qquad e_0=C_0-\sum_{i=1}^g f_i,
\end{equation*}
and $\l_f$ becomes a constant, so we drop the index $f$ from the
notation.

We will use the theory of canonical transformations to show that $B_\xi$
has the spectrality property and we will find along the way an alternative,
simpler, proof that $B_\xi$ is a Poisson map. Recall that a transformation
(bijective map) between ($2g$-dimensional) symplectic manifolds is canonical
(symplectic) if and only if it has a local generating function $F$, i.e.,
in terms of canonical variables $(x_i,y_i)$ and $(\tilde x_i,\tilde y_i)$
one has a function $F(x_1,\dots,x_g\vert\tilde x_1,\dots,\tilde x_g)$ such
that

\begin{equation}\label{gen_prop}
  y_i=\frac{\partial F}{\partial x_i} \qquad\hbox{and}\qquad
      \tilde y_i=-\frac{\partial F}{\partial \tilde x_i}.
\end{equation}
In turn this is equivalent to the compatibility relations
\begin{equation*}
  \frac{\partial y_i}{\partial x_j}= \frac{\partial y_j}{\partial x_i},\qquad
  \frac{\partial \tilde y_i}{\partial \tilde x_j}= \frac{\partial \tilde
  y_j}{\partial \tilde x_i},\qquad
  \frac{\partial \tilde y_i}{\partial x_j}= - \frac{\partial y_j}{\partial \tilde x_i},
\end{equation*}
where $i,j=1,\dots,g$; in these formulas one views the transformation
locally as a map $(x_1,\dots,x_g,\tilde x_1,\dots,\tilde x_g)\to
(y_1,\dots,y_g,\tilde y_1,\dots,\tilde y_g)$. In the present case this
means that we have to view $h_1,\dots,h_g,\tilde h_1,\dots,\tilde h_g$ as
functions of $f_1,\dots,f_g,\tilde f_1,\dots,\tilde f_g$ and that we need
to verify the following compatibility relations
\begin{equation}\label{int_cond}
  f_j\frac{\partial h_i}{\partial f_j}= f_i\frac{\partial h_j}{\partial f_i},\qquad
  \tilde f_j\frac{\partial \tilde h_i}{\partial \tilde f_j}= \tilde f_i\frac{\partial
\tilde
  h_j}{\partial \tilde f_i},\qquad
  f_j\frac{\partial \tilde h_i}{\partial f_j}= - \tilde f_i\frac{\partial h_j}{\partial
\tilde f_i}.
\end{equation}
To do this we need to express the variables $h_i,\,\tilde h_i$ and $\beta$
in terms of the variables $f_i$ and $\tilde f_i$.  Multiplying both sides
of (\ref{ut_odd}) by $\l-x$ and comparing the leading terms in $x$ we find
$\b^2=\l+w_0-u_1$, leading to the following expression for $\b$ as a
function of $\{\tilde f_i,f_i\}_{i=1}^g$:
\begin{equation}\label{beta2}
  \b^2=\l+C_0-\sum_{i=1}^g(\tilde f_i+f_i).
\end{equation}
Excluding the $e$-variables from the equations (\ref{i}) of the map
$B_\xi: \{h_i,f_i\}_{i=1}^g \mapsto \{\tilde h_i,\tilde f_i\}_{i=1}^g$
we find the following $2g$ equations:
\begin{align}
  &(h_i+\b f_i)^2-(\l-a_i)\tilde f_if_i-C_i=0,\label{1}\\
  &\tilde h_i=-h_i+\b(\tilde f_i-f_i).\label{2}
\end{align}
Notice that with $\b$ from (\ref{beta2}) the first equation defines $h_i$
and then the second equation defines $\tilde h_i$, both as implicit
functions of the variables $\{\tilde f_i,f_i\}_{i=1}^g$. Straightforward
computation leads to
\begin{equation*}
  \frac{\partial h_i}{\partial f_j}=\frac{f_i}{2\b}\qquad \hbox{and}\qquad
  \frac{\partial \tilde h_i}{\partial \tilde f_j}=-\frac{\tilde f_i}{2\b}
\end{equation*}
for $i\neq j$ and to
\begin{equation*}
  \frac{\partial h_i}{\partial \tilde f_j}=\frac{f_i}{2\b}+\frac{(\l-a_i)f_i}{2(h_i+\b
f_i)}\delta_{ij}
  \qquad\hbox{and}\qquad
  \frac{\partial \tilde h_i}{\partial f_j}=-\frac{\tilde f_i}{2\b}-\frac{(\l-a_i)\tilde
f_i}{2(h_i+\b f_i)}\delta_{ij},
\end{equation*}
for any $i,j$. The compatibility conditions (\ref{int_cond}) follow at once.

In fact, in the same way we can prove another property of the \BT, its
spectrality, which means that the variables $\hat \mu$ and $\l$ are also
canonical, in a sense, or more precisely, that the parameter $\l$ enters in
the generating function $F=F_\l$ in such a way that for the $\hat\mu$ being
expressed in terms of $\{\tilde f_i,f_i\}_{i=1}^g$ variables we have a
similar expression as in (\ref{gen_prop}):
$$
\hat \mu=\frac{\partial F_\l}{\partial \l}\,.
$$
It follows that the following compatibility conditions are sufficient for
proving the spectrality property of the \BT:
\begin{equation}\label{comp_mu}
  f_i\,\frac{\partial \hat\mu}{\partial f_i} = \frac{\partial h_i}{\partial
  \l}\qquad\hbox{and}\qquad
  \tilde f_i\,\frac{\partial \hat\mu}{\partial \tilde f_i} =-\frac{\partial
  \tilde h_i}{\partial \l}\,.
\end{equation}
It is easily checked from (\ref{3}) that these compatibility conditions
indeed hold; the values of the two expressions in (\ref{comp_mu}) are given
by
\begin{equation*}
  -\frac{f_i}{2\b}+\frac{f_i\tilde f_i}{2(h_i+\b f_i)}\qquad\hbox{and}\qquad
  -\frac{\tilde f_i}{2\b}+\frac{f_i\tilde f_i}{2(h_i+\b f_i)}.
\end{equation*}
We have shown that our \BT s are Poisson maps and have the spectrality
property when $\varphi$ is monic of degree $g$ and is irreducible. Obviously
the fact that $\varphi$ is monic is inessential. Moreover, all Poisson
brackets are polynomial in terms of the roots $a_i$ of $\varphi$ hence these
properties hold when $\varphi$ is any polynomial of degree at most $g$.

  \subsection{Addition formulas for the $\wp$ function} %

\label{wp_section}
In this paragraph we show that our formulas (\ref{ut_odd}) and
(\ref{vt_odd}) generalize the classical addition formulas for the
Weierstra\ss\ $\wp$ function to the case of (families of)
hyperelliptic curves. Let $\Cur$ be an elliptic curve, written in the
Weierstra\ss\ form
\begin{equation*}
  Y^2=4X^3-g_2X-g_3.
\end{equation*}
Points on this curve are parametrized by $\wp$ and its derivative
$\wp'$: for any $(X,Y)\in \Cur$ there is a $z\in\C$ such that
$(X,Y)=(\wp(z),\wp'(z))$. We write the equation of $\Cur$ as
$y^2=f(x)=x^3-(g_2/4)x-(g_3/4)$, thereby fixing $f\in\Px_3$. We take
two generic points on $\Cur$ and their sum ($\Cur$ is its own
Jacobian, hence a group): $(\l_f,\mu_f)+(p,q)=(\pt,\qt)$. On the one hand
we can associate to the points $(p,q)$ and $(\pt,\qt)$ the
corresponding polynomials of the Mumford system, on the other hand we
can write them in terms of the $\wp$ function. As for the former we
get
\begin{align*}
  u(x)&=x+u_1=x-p,\\
  v(x)&=v_1=q,\\
  w(x)&=x^2-u_1x+w_1=x^2+px+(4p^2+g_2)/4,
\end{align*}
for $(p,q)$ and we get similar formulas for $(\pt,\qt)$ by putting
tildes over all variables. In terms of $p,q,\pt$ and $\qt$ formulas
(\ref{ut_odd}), (\ref{vt_odd}) and (\ref{b_odd}) (in that order) take the
form
\begin{equation}\label{befs}
  \b^2=p+\pt+\l,\qquad \b =-\frac{q+\qt}{p-\pt}=\frac{\mu-q}{\l_f-p}.
\end{equation}
As for the latter, let
$(p,q)=(\wp(z),\wp'(z)/2),\,(\pt,\qt)=(\wp(\zt),\wp'(\zt)/2)$ and
$(\l_f,\mu_f)=(\wp(z'),\wp'(z')/2)$. Then (\ref{befs}) reduces, after
eliminating $\b $ to the following classical formulas:
\begin{align*}
  {\frac14\left(\frac{\wp'(z)+\wp'(\zt)}{\wp(z)-\wp(\zt)}\right)^2}
       &=\wp(z)+\wp(\zt)+\wp(z'),\\
  {\frac14\left(\frac{\wp'(z')-\wp'(z)}{\wp(z')-\wp(z)}\right)^2}
       &=\wp(z)+\wp(\zt)+\wp(z'),
\end{align*}
with $\zt=z+z'.$

\goodbreak

\section{\BT s in more complex situations} 
%
\label{complex}

  \subsection{The even Mumford system} 
%
\label{even_ss}
\def\Pg{{\Px_{2g+2}}}
The Mumford system has a twin which was introduced by the second author in
\cite{Van3}, where it was called the even master system; in this text we will call it
the {\sl even\/} Mumford system. The phase space $M_g$ of the even Mumford system
consists of Lax operators
\begin{equation*}
  L(x)=\mat{v(x)}{w(x)}{u(x)}{-v(x)},
  \label{Laxmatrix_even}
\end{equation*}
where $u(x),\,v(x)$ and $w(x)$ are now subject to the following
constraints: $u(x)$ and $w(x)$ are monic and their degrees are
respectively $g$ and $g+2$; the degree of $v(x)$ is at most $g-1$. In
this case we write
\begin{align*}
  u(x)&=x^g+u_1x^{g-1}+\cdots+u_g, \\
  v(x)&=v_1x^{g-1}+\cdots+v_g, \\
  w(x)&=x^{g+2}+w_{-1}x^{g+1}+\cdots+w_g.
\end{align*}
The map $\chi:M_g\to\Pg$ is defined as in (\ref{mmap}); notice that
$\chi$ takes its values now in the affine space of monic polynomials
of degree $2g+2$, explaining the adjective {\sl even}. The main
difference between the even and the odd Mumford system is that the
spectral curves $\Cur_f:y^2=f(x)=u(x)w(x)+v^2(x)$ have now two points
at infinity, a fact which has drastic consequences for the geometry of
the integrable system (see \cite{Van1}).

Let us first construct \BT s for this system by using the approach described in
Paragraph \ref{back}. We denote by $\Ug$ the universal curve over $\Pg$ and we
consider sections of the natural projection $\pi:\Ug\to\Pg$, as in Paragraph
\ref{back}. In this case there is no natural section of $\pi:\Ugb\to\Pg$, so we need
to choose two sections of $\pi$ to construct a \BT\ (for the existence of such
sections the remarks from Paragraph \ref{existence_parg} apply). To simplify the
formulas for the \BT\ and to make them very similar to the formulas in the odd case
we pick one of the sections such that every $f\in\Pg$ gets mapped to one of the two
points at infinity, i.e. in $\Cb_f\setminus \Cur_f$. We denote this section by
$\xi_\infty$ and we pick another section $\xi$. Since Mumford's prescription
(\ref{uvtodiv1}) and (\ref{uvtodiv2}) applies unchanged, the following variant to
(\ref{F_odd}) realizes the linear equivalence which is needed in order to express a
shift over $[\xi(f)-\xi_\infty(f)]$ on $\Jac(\Cb_f)$,
\begin{align}
  F(x,y)&=\frac{y+v(x)+u(x)(\pm (x-\l_f)+\b)}{u(x)(x-\l_f)}
  \label{F_even}\\
  &=\frac{y+v(x)+\b u(x)}{u(x) (x-\l_f)}\pm1\,,\notag
\end{align}
where $\b $ is such that the numerator vanishes at $(\l_f,-\mu_f)$, so that
\begin{equation}\label{b_even}
  \b =\frac{\mu_f-v(\l_f)}{u(\l_f)}\,.
\end{equation}
The $\pm$ in (\ref{F_even}) depends on the chosen section
$\xi_\infty$, its actual value, for a given $f$ being determined by
expressing $x$ and $y$ in terms of a local parameter at the point
$\xi_\infty(f)$. The rest of the computation is similar to the one in
Paragraph \ref{back}, giving
\begin{align}
  \ut(x)&=\frac{u(x)(x-\l_f\pm\b)^2\pm2v(x)(x-\l_f\pm\b)-w(x)}
           {(u_1-w_{-1}-2\l_f \pm 2\b )(x-\l_f)}\,,\notag\\
  \vt(x)&=-v(x)\mp u(x)(x-\l_f\pm\b)\pm\ut(x)(x-\l_f+u_1-\ut_1\pm\b)\,,
\label{BT_even}\\
  \wt(x)&=(u(x)w(x)+v^2(x)-\vt^2(x))/\ut(x)\,,\notag\\
  \b &=\frac{\mu_f-v(\l_f)}{u(\l_f)}\,.\notag
\end{align}
The value of the variable $\ut_1$ in terms of the original variables
is computed from the first equation in (\ref{BT_even}) to be given by
\begin{equation*}
  \ut_1=\lambda_f+\frac{u_2\pm 2v_1-w_0\pm 2u_1(\b\mp\l_f)+(\b\mp\l_f)^2}
    {u_1-w_{-1}-2\l_f\pm 2\b}\,.
\end{equation*}
The matrix $M(x)$, defined as in (\ref{Lt}) can in this case be taken as
\begin{equation}\label{M_even}
  \mat{x-\l_f+u_1-\ut_1\pm\b}
      {\b(u_1-\ut_1\pm\b)\pm(x-\l_f)(x+\l_f+w_{-1}-\ut_1)}
      {\pm1}
      {x-\l_f\pm\b }.
\end{equation}
Notice that $\det M(x) = (x-\l_f)(u_1-w_{-1}-2\l_f\pm 2\b)$.

The integrable vector fields of the even Mumford system are Hamiltonian
with respect to a family of Poisson brackets, similar to the brackets
(\ref{brackets}): if $\varphi$ is a univariate polynomial of degree at most
$g$ then one finds precisely the brackets (\ref{brackets}), except for the
following two brackets
\begin{align*}
  \{v(x),w(y)\}^\varphi&=\frac{1}{x-y}\,\left(w(x)\varphi(y)-
           w(y)\varphi(x)\right)-
            \alpha(x,y) u(x)\varphi(y), \\
  \{w(x),w(y)\}&=2\alpha(x,y)\left(v(x)\varphi(y)-v(y)\varphi(x)\right), \\
  \alpha(x,y)&=x+y+w_{-1}-u_1,
\end{align*}
define a Poisson structure on $M_g$.  Assuming $\varphi(x)$ monic and irreducible,
$\varphi(x)=(x-a_1)(x-a_2)\cdots(x-a_g)$, we define an affine map
$M_g\to\C^{3g+2}$ by
\begin{equation*}
  \left(\frac{u(x)}{\varphi(x)},\frac{v(x)}{\varphi(x)},\frac{w(x)}{\varphi(x)}\right)
    =\left(1+\si\frac{f_i}{x-a_i},\si\frac{h_i}{x-a_i},
  x^2+e_{-1}x+e_0+\si\frac{e_i}{x-a_i}\right).
\end{equation*}
As in the case of the Mumford system we find that the variables
$\{h_i,e_i,f_i\}_{i=1}^g$ are generators for the direct sum of $g$ copies
of the Lie-Poisson algebra of $\sllie(2)$.
The equation of the spectral curve takes the form
\begin{equation*}
  \frac{y^2}{\varphi^2(x)}=\frac{f(x)}{\varphi^2(x)}=
  x^2+C_{-1}x+C_0+\sum_{i=1}^g\left(\frac{C_i}{(x-a_i)^2}+\frac{H_i}{x-a_i}\right),
\end{equation*}
where $C_i=h_i^2+e_if_i$, the Casimir element coming from the $i$-th copy
of $\sllie(2)$; moreover $C_{-1}=e_{-1}+\si f_i$ and $C_0=e_0+\si
f_i(C_{-1}+a_i) -(\si f_i)^2$ are extra Casimirs. Fixing a generic
symplectic leaf, these Casimirs are used to eliminate the variables
$e_{-1},\dots,e_{g}$ giving the following equations for the map
$(i=1,\dots,g)$
\begin{align*}
  &\left(\si 2f_i-2\l\pm2\b-C_{-1}\right)(\l-a_i)f_i\tilde f_i+(f_i(a_i-\l\pm\b)\pm
h_i)^2-C_i=0,\\
  &\tilde h_i=-h_i\mp(f_i-\tilde f_i)(a_i-\l\pm\b)\pm \tilde f_i\si(f_j-\tilde f_j)
\end{align*}
and the following equation for $\b$
\begin{equation*}
  \b^2\pm2(u_1-\tilde u_1)\b -\l^2+\l(2\tilde u_1-w_{-1}-u_1)-u_1\tilde
    u_1-w_0+u_2+\tilde u_1w_{-1}\pm2v_1=0,
\end{equation*}
where
\begin{align*}
  u_1&=\si(f_i-a_i),\qquad\qquad\qquad\,\, v_1=\si h_i\\
  u_2&=\sum_{i<j}a_ia_j-\sum_{i\neq j}a_if_j\qquad\qquad
w_{-1}=C_{-1}-\si(a_i+f_i)\\
  w_0&=C_0-C_{-1}\si(a_i+f_i)+\left(\si f_i\right)^2+\sum_{i<j}a_ia_j
+\sum_{i\neq
j}a_if_j.
\end{align*}
Using these formulas the verification of (\ref{int_cond}) and (\ref{comp_mu})
(where
$\hat\mu_f$ is in this case again
defined by $\hat\mu_f=\mu_f/\varphi(\l_f)$ and it is assumed that $\l_f$
depends on the
Casimirs only) is now
straightforward (but lenghty). This shows again that our maps $B_\xi$
are Poisson maps and
have the spectrality property
when $\l_f$ depends on the Casimirs of $\Pb^\varphi$ only.

In order to show that our maps $B_\xi$ give a discretization of the even
Mumford system,
we proceed as in Paragraph
\ref{cont}. We let $\l_f=1/t$ so that the first few terms of $\b $ are given by
\begin{equation*}
  \b =\mp\frac1t\left(1+\frac{w_{-1}-u_1}2t
+\frac18(3u_1^2-2u_1w_{-1}-w_{-1}^2-
        4u_2+4w_0\pm 8v_1)t^2+O(t^3)\right)
\end{equation*}
A direct substitution in (\ref{BT_even}) yields
\begin{align*}
  \ut(x)&=u(x)\mp v(x)t+O(t^2)\,,\\
  \vt(x)&=v(x)\mp\frac12(-w(x)+u(x)(x^2+(w_{-1}-u_1)x+\\
        &\qquad\qquad\qquad\qquad\qquad u_1^2+w_0-u_2-u_1w_{-1}))
          t+O(t^2)\,,\\
  \wt(x)&=w(x)\pm v(x)(x^2+(w_{-1}-u_1)x+u_1^2+w_0-u_2-u_1w_{-1})t+O(t^2)\,.
\end{align*}
Moreover we can construct the analogs of Mumford's vector fields
$X_a$. We proceed as in Paragraph \ref{cont}, but special care has to
be taken because now the curve has two points at infinity, namely
$\infty_f$ and the point that corresponds to $\infty_f$ under the
hyperelliptic involution; the latter point will be denoted by
$\infty'_f$. Fixing a section $\xi$, we write $\xi(f)=(a_f,b_f)$ and
we do a translation over $[(a_f,-b_f)-\infty_f]$. The matrix going with
this transformations is denoted by $P(x)$. Then we translate over
$[(\l_f(t),\mu_f(t))-\infty'_f]$; its matrix is denoted by
$Q_t(x)$. The product then corresponds to a translation over
$[(\l_f(t),\mu_f(t))-(a_f,b_f)]$. Explicitly, for $P(x)$ we take the
lower signs in (\ref{M_even}) to get
\begin{equation*}
  P(x)=\mat{x-a+u_1-\ut_1+\b}{\b(\ut_1-u_1-\beta)-(x-a)(x+a+w_{-1}-\ut_1)}
           {-1}{x-a+\b}
\end{equation*}
with
\begin{align*}
  \ut_1&=a_f+\frac{u_2-2v_1-w_0-2u_1(a_f-\b)+(a_f-\b)^2}
{u_1-w_{-1}-2a_f+2\b},\cr
  \b &=\frac{b_f+v(a_f)}{u(a_f)}=\frac{w(a_f)}{b_f-v(a_f)}.
\end{align*}
For $Q_t(x)$ we take the upper sign and we find
\begin{equation*}
  Q_t(x)=\mat{x-\l_f(t)+\ut_1-\utt_1+\b(t)}
             {\star}
             1
             {x-\l_f(t)+\b(t)}
\end{equation*}
where
$\star={\b(t)(\ut_1-\utt_1+\b(t))+(x-\l_f(t))(x+\l_f(t)+\wt_{-1}-\utt_1)}$
and
\begin{align*}
  \utt_1&=\l_f(t)+\frac{\ut_2+2\vt_1-\wt_0+2\ut_1(\b(t)-\l_f(t))+(\b(t)+\l_f(t))^2}
                      {\ut_1-\wt_{-1}-2\l_f(t)+2\b(t)},\cr
  \b (t)&=\frac{\mu_f(t)-\vt(\l_f(t))}{\ut(\l_f(t))}.
\end{align*}
In order to express $\utt_1$ in terms of the original phase variables,
as needed in the computation, one needs explicit formulas for
$\ut_2,\vt_1,\wt_{-1}$ and $\wt_0$. For $\ut_2$ and $\vt_1$ we find by
expanding the first \BT\ in terms powers of $t$
\begin{align*}
  \ut_2&=a\ut_1+\frac{u_3-2(a-\b)u_2+(a-\b)^2u_1-2v_2+2(a-\b)v_1-w_1}
                     {u_1-w_{-1}-2a_f+2\b},\cr
  \vt_1&=-v_1+u_2-(a-\b)u_1-\ut_2+\ut_1(a-u_1+\ut_1-\b).
\end{align*}
We find as in the case of the Mumford system that $\b(0)=u_1-\ut_1+\b$
and that
\begin{equation*}
  \b'(0)=1-(u_1-w_{-1}-2a+2\b)\frac{u(a)}{2b}.
\end{equation*}
As we have seen in the Mumford case the vector field which corresponds
to the deformation family is given by
\begin{equation*}
  \frac{dL}{dt_{a_f}}(x)=\left[Q'_0(x)\, Q_0^{-1}(x),L(x)\right],
\end{equation*}
which leads by direct substitution to
\begin{equation*}
  \frac{dL}{dt_{a_f}}(x)=\frac1{2b_f}\left[\frac{L(a_f)}{x-a_f}+
         \mat0{u(a_f)(x+a_f+u_1-w_{-1})}00,L(x)\right].
\end{equation*}
As far as we could check these vector fields are new.
%
  \subsection{Generalized Jacobians (odd case)} 

\label{gen_par}
\def\Pg{{\Px_{2g+1}}}
We now consider a first case in which the fibers of the moment map
are affine parts of generalized (hyperelliptic) Jacobians. The main
difference between the generalized Jacobian case and the usual case is
that generalized Jacobians have a larger symmetry group, leading to
more general \BT s.

We first define phase space, which is denoted by $\Mh_g$, a moment map
$\hat\chi:\Mh_g\to \Pg$, we construct a natural map $\pi:\Mh_g\to M_g$
onto the phase space of the Mumford system, and we give a geometric
description of the fibers of $\chi$. For any $g\geq1,$ $\Mh_g$ is the
space of all Lax matrices of the form
\begin{equation*}
  L(x)=\mat{V(x)}{W(x)}{U(x)}{-V(x)},
  \label{Laxmatrix_gen}
\end{equation*}
where the entries of $L(x)$ are now subject to the following
constraints: $U(x)$ and $W(x)$ are monic and their degrees are
respectively $g$ and $g+1$; the degree of $V(x)$ is at most
$g$. Writing
\begin{align*}
  U(x)&=x^g+U_1x^{g-1}+\ldots+U_g, \\
  V(x)&=V_0x^{g}+\ldots+V_g, \\
  W(x)&=x^{g+1}+W_0x^{g}+\ldots+W_g,
\end{align*}
we take the coefficients of these three polynomials as coordinates on
$\Mh_g$. It is clear that the group of matrices of the form
\begin{equation}\label{upper}
  N_\tau=\mat1{-\tau}01
\end{equation}
acts on $\Mh_g$ by the adjoint action, where $\tau$ is any function on
$\Mh_g$. In particular, taking $\tau=V_0$ we get a map onto a subspace
which is exactly the phase space $M_g$ of the Mumford system; we
denote this natural map by $\pi$ and denote the composition
$\chi\circ\pi$ by $\hat\chi$; explicitly $\hat\chi$ is given by
$L(x)\mapsto -\det L(x)$. For $f\in\Pg$ such that $\Cur_f$ is smooth
the fiber $\chi^{-1}(f)$ is an affine part of $\Sym^{g+1}\Cb_f$, the
$(g+1)$-th symmetric product of $\Cb_f$ (recall that $\Cb_f$ has genus
$g$). To see this, one associates to
$(U(x),V(x),W(x))\in\chi^{-1}(f)$ the divisor
$D=\sum_{i=1}^{g+1}(x_i,y_i)$, where $x_i$ are the roots of $W(x)$ and
$y_i=-V(x_i)$. It is easy to show that this realizes a bijection
between $\chi^{-1}(f)$ and an affine part of $\Sym^{g+1}(\Cb_f)$
\footnote{From this  description it follows easily that the fiber
$\chi^{-1}(f)$ can also be described as an affine part of the
generalized Jacobian of $\Cur_f$ with respect to the divisor
$2\infty_f$. See \cite{Ser}.}. The rational function
\begin{equation*}
  \frac{y-V(x)}{W(x)}=\frac{U(x)}{y+V(x)}
\end{equation*}
shows that $\D$ is linearly equivalent to the divisor
$\D'+\infty_f=\sum_{i=1}^g (x'_i,y'_i)+\infty_f$, where $x'_i$ are the
zeros of $U(x)$ and $V(x'_i)=y'_i$ for $i=1,\dots,g$. This gives a
geometric interpretation of the map $\pi$, and it shows that, under
the above correspondence between points of $\Mh_g$ and divisors, the
adjoint action by $N_\tau$ maps divisors to linearly equivalent
divisors.

We will show that this geometric picture leads, via our geometric
construction of \BT s, to a family of \BT s
$B_{\xi,\a}:\Mh_g\to\Mh_g$ which makes the following diagram
commutative. \par
\begin{equation}\label{diag}
  \begin{diagram}
    \node{\Mh_g}\arrow{e,t,..}{B_{\xi,\a}}
                \arrow{s,l}{\pi}
    \node{\Mh_g}\arrow{s,l}{\pi}\\
    \node{M_g}\arrow{e,t}{B_{\xi}}
    \node{M_g}
  \end{diagram}
\end{equation}
It should be clear that, since we are forced to work with divisors, we
cannot write (\ref{BT_gen}) as a \emph{definition} for $B_{\xi,\a}$
because the effective divisor of degree $g+1$ that corresponds to a
line bundle of degree $g+1$ is not unique. Accordingly we write down a
general formula for a map satisfying (\ref{BT_gen}) and then we
specialize the arbitrary function that figures in it so as to obtain a
\BT.  Explicitly, we let $\xi(f)=(\l_f,\mu_f,f)$, as before, and we
consider for a generic point $\UVW\in\Mh_g$ the following function
\begin{equation*}
  F(x,y)=\frac{(y-V(x))(x-\l_f+\a\b)+\a W(x)}{W(x)(x-\l_f)}.
\end{equation*}
We have chosen a combination of the parameters $\a$ and $\b$ such
that, when we express that the numerator of $F$ vanishes at
$(\l_f,-\mu_f)$, then we find
\begin{equation*}
  \b=\frac{W(\l_f)}{\mu_f+V(\l_f)}=\frac{\mu_f-V(\l_f)}{U(\l_f)},
\end{equation*}
so that $\b$ is formally given by the same formula (\ref{b_odd}) as in
the Mumford system. With this choice of $\beta$ we find for any $\a$
that $F(x,y)$ has
$\D+(\l_f,\mu_f)=\sum_{i=1}^{g+1}(x_i,y_i)+(\l_f,\mu_f)$ as its polar
divisor and vanishes at infinity. It follows that the other zeros of
$F(x,y)$ give a divisor $\Dt$ which is linearly equivalent to the
divisor $\D$ which is associated to $\UVW$, up to a shift over
$(\l_f,\mu_f)-\infty_f$.  Multiplying $F(x,y)$ by
$(y+V(x))(x-\l_f+\a\b)-\a W(x)$ and using $y^2=U(x)W(x)+V^2(x)$ we
find an equation for the $x$-coordinates of the image divisor and we
deduce, as in the case of the Mumford system,
\begin{equation}\label{W}
  \tilde W(x)=-\frac{(x-\l_f+\a\b)^2U(x)+2\a(x-\l_f+\a\b)V(x)-\a^2W(x)}{\l_f-x}.
\end{equation}
By interpolation at the zeros of $\Wt$ we also find
\begin{equation}\label{V}
  \tilde V(x)=\frac{\b(x-\l_f+\a\b)U(x)+(x-\l_f+2\a\b)V(x)-\a W(x)}{\l_f-x},
\end{equation}
and the formula for $\tilde U(x)$ follows from $\tilde U(x)\tilde
W(x)+\tilde V^2(x)= U(x)W(x)+V^2(x)$,
\begin{equation}\label{U}
  \tilde U(x)=\frac{\b^2U(x)+2\b V(x)-W(x)}{\l_f-x}.
\end{equation}
This gives explicit formulas for the map $B_{\xi,\a}$. In terms of
matrices, $B_{\xi,\a}$ is given by $L\mapsto M LM^{-1}$, where $M$ can
be taken as follows:
\begin{equation}\label{M_gen}
  M(x)=\mat{\a}{x-\l_f+\a\b}1{\b}.
\end{equation}
The commutativity of (\ref{diag}) is a direct consequence of the
equality $N_{-V_0+\a-\b}M=\bar M N_{V_0}$, where $\bar M$ is given by
\begin{equation*}
  \bar M(x)=\mat{\b+V_0}{x-\l_f+(\b+V_0)^2}1{\b+V_0}.
\end{equation*}
If we compare (\ref{M_odd}) and (\ref{M_gen}) then we see that both
matrices coincide when $\a=\b$, but, as we will see, the choice
$\a=\b$ does not lead to a \BT\ (when $\a=\b$ it is not a Poisson map).
\par
We now come to poissonicity of the maps that we have constructed. The
Poisson structure of the generalized Mumford system is given, in the
notation of Paragraph \ref{odd_poisson_section}, by
\begin{equation}\label{PB_gen_len}
  \PB{L(x)}{L(y)}= \left[r(x-y),L_1(x)\varphi(y)+\varphi(x)L_2(y)\right],
\end{equation}
where $\varphi(x)$ is a polynomial of at most degree $g$. We take
$\l_f$ to be dependent on the Casimirs only and we compute, as before,
the brackets with $\beta$, giving
\begin{align}
  \{U(x),\b\}^\varphi&=\frac{\mu_f\varphi(x)-\varphi(\l_f)(V(x)+\b
U(x))}{\mu_f(x-\l_f)},\notag\\
  \{V(x),\b\}^\varphi&=-\frac{2\mu_f\b\varphi(x)-\varphi(\l_f)(\b^2U(x)+W(x))}
                              {2\mu_f(x-\l_f)},\label{bra_b}\\
  \{W(x),\b\}^\varphi&=-\b\frac{\b\mu_f\varphi(x)+\varphi(\l_f)(\b
V(x)-W(x))}{\mu_f(x-\l_f)}.\notag
\end{align}
Using these formulas we can determine for which choices of $\a$ (which
could, a priori, be any function on phase space) the map
$\UVW\to\UVWt$ is a Poisson map. A (quite long) computation leads to
the following conditions on $\a$.
\begin{align*}
\{\alpha,U(x)\}&=-C\frac{V(x)+\beta U(x)}{x-\l_f},\\
\{\alpha,V(x)\}&=-C\frac{W(x)+\beta^2 U(x)}{2(x-\l_f)}+D,\\
\{\alpha,W(x)\}&= C\beta\frac{W(x)-\beta V(x)}{x-\l_f}+\varphi(x),\\
\{\alpha,\beta\}&=\varphi(\l_f)/(2\mu_f).
\end{align*}
In these formulas $C$ and $D$ are any functions on phase
space. However, since the left hand side of the first three
expressions is polynomial in $x$, the same must be true for the right
hand side, which implies that $C=0$. Using the last equation and the
definition of $\beta$ we find that $D=0$ and we are left with
\begin{align}
\{\alpha,U(x)\}&=\{\alpha,V(x)\}=0,\notag\\
\{\alpha,W(x)\}&=\varphi(x),\label{bra_a}\\
\{\alpha,\beta\}&=\varphi(\l_f)/(2\mu_f).\notag
\end{align}
It turns out that there is such an $\alpha$, namely $\a=V_0$; to
obtain the most general solution it suffices to add any Casimir of
$\varphi$ to $V_0$. A direct check that one gets for those values of
$\a$ indeed a Poisson map can be done quite easily by using the
following formulas, which follow from (\ref{PB_gen_len}),
(\ref{bra_b}) and (\ref{bra_a}).
\begin{align*}
  \PB{L(x)}{M(y)}&=
      \left( \frac{\varphi(\l_f)}{2\mu_f} \left[L(x),N(x)\right]
            +\varphi(x)N(x)\right)
      \t\frac{\partial M}{\partial\b}
      -\varphi(x)\frac{\partial^2 M}{\partial\a\partial\b}\t
         \frac{\partial M}{\partial\a},\\
  \PB{M(x)}{L(y)}&=-
      \frac{\partial M}{\partial\b}\t
      \left( \frac{\varphi(\l_f)}{2\mu_f} \left[L(y),N(y)\right]
            +\varphi(y)N(y)\right)
      +\varphi(y)\frac{\partial M}{\partial\a}\t
            \frac{\partial^2 M}{\partial\a\partial\b},\\
  \PB{M(x)}{M(y)}&=
   -\frac{\varphi(\l_f)}{2\mu_f}
     \left(\begin{array}{cccc}
           0&\a&-\a&0\\
           0&1&0&\b\\
           0&0&-1&-\b\\
           0&0&0&0
           \end{array}\right)
\end{align*}
where $N(x)=\frac1{\l_f-x}\mat\b{\b^2}{-1}{-\b}$.
In conclusion we have shown that when $\l_f$ and $\a-V_0$ depend only
on the Casimirs then the map $B_{\xi,\a}$ is a \BT\ for the
generalized Mumford system.

In order to check spectrality of the map $B_{\xi,\a}$ when $\l_f$ and
$\a-V_0$ depend only on the Casimirs one proceeds as in the case of the
Mumford system. We fix a monic polynomial $\varphi(x)$ of degree $g$ with
distinct roots $a_1,\dots,a_g$ and we define an affine map
$\Mh_g\to\C^{3g+2}$ by
\begin{equation}\label{gen_gaud}
  \left(\frac{U(x)}{\varphi(x)},\frac{V(x)}{\varphi(x)},\frac{W(x)}
{\varphi(x)}\right)
         =\left(1+\si\frac{f_i}{x-a_i},
  h_0+\si\frac{h_i}{x-a_i},x+f_0+\si\frac{e_i}{x-a_i}\right).
\end{equation}
In this case we get the brackets (\ref{sl2brackets}) with in addition one
non-trivial bracket, $\{h_0,f_0\}=1$.  We denote the Casimir element coming
from the $i$-th copy of $\sllie(2)$ by $C_i$, $C_i=h_i^2+e_if_i$ and we
denote the Casimir $\a-V_0$ by $C$. We fix a
symplectic leaf and we express the variables $f_0,\dots,f_g,\tilde
f_0,\dots,\tilde f_g$ in terms of $h_0,\dots,h_g,\tilde h_0,,\dots,\tilde h_g$
and $\l$. To do this, first notice that
\begin{equation*}
  \a=h_0+C, \qquad \mbox{and} \quad \b=C-\tilde h_0,
\end{equation*}
as follows easily from (\ref{gen_gaud}) and (\ref{V}).
The formulas for the variables $f_1,\dots,f_g$ follow from
\begin{align}\label{sim_eq}
  &(\b f_i+h_i)^2-f_i\tilde f_i(\l-a_i)-C_i=0,\\
  &\tilde h_i+h_i-\a \tilde f_i+\b f_i=0,
\end{align}
which one derives from the equations (\ref{W}), (\ref{V}) and (\ref{U}) for
$B_\xi$, expressed in terms of the variables $f_i$ and $h_i$. Indeed, if we
use the second equation to eliminate $\tilde f_i$ from the first equation
we get
\begin{equation}
  f_i^2\tilde  h_0(a_i-\lambda-h_0\tilde h_0)+f_i((\lambda-a_i)(\tilde
  h_i+h_i)+2h_ih_0 \tilde h_0)+h_0(C_i-h_i^2)=0,
\end{equation}
and this defines $f_1,\dots,f_g$ as a function of the variables $h_j$ and
$\tilde h_j$; the second equation in (\ref{sim_eq}) then defines $\tilde
f_1,\dots,\tilde f_g$ as a function of these variables. As for $f_0$ and
$\tilde f_0$, they are given by
\begin{align*}
  &f_0=-\l+\si \tilde f_i+\tilde h_0^2-2h_0\tilde h_0,\\
  &\tilde f_0=-\l+\si f_i+h_0^2-2 h_0\tilde h_0,
\end{align*}
as follows also from (\ref{W}), (\ref{V}) and (\ref{U}). Using these
formulas it is straightforward to verify the following integrability
conditions ($i,j=1,\dots,g$)
\begin{align*}
  \pp{f_i}\l&=-f_i\pp{\hat\mu}{h_i}=-\frac{\a f_i\tilde f_i}{(\l-a_i)(\b
  f_i+\a\tilde f_i)-2\a\b(h_i+\b f_i)},\\
  \pp{\tilde f_i}\l&=\tilde f_i\pp{\hat\mu}{h_i}=-\frac{\b f_i\tilde f_i}
{(\l-a_i)(\b
  f_i+\a\tilde f_i)-2\a\b(h_i+\b f_i)},\\
  \pp{f_0}\l&=-\pp{\hat\mu}{h_0}=-1-\b\si \frac{f_i\tilde f_i}{(\l-a_i)(\b
  f_i+\a\tilde f_i)-2\a\b(h_i+\b f_i)},\\
  \pp{\tilde f_0}\l&=\pp{\hat\mu}{\tilde h_0}=-1-\a\si \frac{f_i\tilde f_i}
{(\l-a_i)(\b
  f_i+\a\tilde f_i)-2\a\b(h_i+\b f_i)}.\\
\end{align*}
This shows that the maps $B_\xi$ have the spectrality property.
In the same
way one can verify the compatibility conditions
\begin{equation*}
  f_j\frac{\partial f_i}{\partial h_j}= f_i\frac{\partial f_j}{\partial h_i},\qquad
  \tilde f_j\frac{\partial \tilde f_i}{\partial \tilde h_j}= \tilde f_i\frac{\partial
\tilde
  f_j}{\partial \tilde h_i},\qquad
  f_j\frac{\partial \tilde f_i}{\partial h_j}= - \tilde f_i\frac{\partial f_j}{\partial
\tilde h_i},
\end{equation*}
giving an alternative proof that the maps $B_\xi$ are Poisson maps.

We now show that these \BT\ discretize the underlying integrable
system. The computation is similar as in the previous cases, except
that one has to choose the Casimir $\a-V_0$ carefully so as to obtain
the identity transformation in the limit $\l_f\to\infty$. Since the
point at infinity of the curve is a Weierstrass point we let $\l=t^{-2}$
and we choose $\alpha=V_0+1/t$. Then
\begin{equation*}
  \b = \frac1t-V_0+\frac12(W_0-U_1+V_0^2)t+O(t^2),
\end{equation*}
and we find by direct substitution
\begin{align*}
  \Ut(x)&=U(x)+2t(V(x)-V_0U(x))+O(t^2),\notag\\
  \Vt(x)&=V(x)+t(U(x)(2x+W_0-U_1-V_0^2)/2-W(x)))+O(t^2),\\
  \Wt(x)&=W(x)-t(V(x)(2x+W_0-U_1-V_0^2)-2V_0W(x))+O(t^2),\notag
\end{align*}
from which we can read off the vector field. For the vector fields
$X_a$ the computation is very similar to the one in the case of the
Mumford system. Namely we take
\begin{equation*}
  P(x)=\mat{\a}{x-a_f-\a\b}{1}{-\b }
\end{equation*}
with $\a=V_0$ and $\b=\frac{b_f+v(a_f)}{u(a_f)}$; moreover we take
\begin{equation*}
  Q_t(x)=\mat{\a(t)}{x-\l_f(t)+\a\b(t)}{1}{\b (t)}
\end{equation*}
where $\a(t)=\tilde V_0=\beta$ (so that in fact $\a$ is independent of
$t$) and $\b(t)=\frac{\mu_f(t)-\Vt(\l_f(t))}{\Ut(\l_f(t))},$ so that
$\b(0)=-\a$. Using $\b'(0)=U(a)/(2b_f)$ we find
\begin{equation*}
  Q_0'(x)Q_0^{-1}(x)=\frac1{2b_f(x-a)}\mat{V(a)-b_f}{W(a)}{U(a)}{-V(a)},
\end{equation*}
so that, after removal of a diagonal matrix, we find the following Lax
equation
\begin{equation*}
  \frac{dL}{dt_{a_f}}(x)=\frac1{2b_f}\left[\frac{L(a_f)}{x-a_f},L(x)\right].
\end{equation*}

We shortly indicate how the above maps $B_{\xi,\a}$ can also be found
from the eigenvectors of the Lax operator. Taking $\va_0=(1,0)$ and
$\va=(\gamma,\delta-x)$ we express that the solutions to
\begin{equation}\label{tilde_eq_1}
  \row10\mat{-\vt(x)-y}{-\wt(x)}{-\ut(x)}{\vt(x)-y}=0,
\end{equation}
are the same as the solutions of
\begin{equation}\label{tilde_eq_2}
  \row{\gamma}{\delta-x}\mat{-v(x)-y}{-w(x)}{-u(x)}{v(x)-y}=0,
\end{equation}
except that (\ref{tilde_eq_2}) also has $(\l_f,-\mu_f)$ as a
solution. By eliminating $y$ from (\ref{tilde_eq_2}) we find that
\begin{equation*}
  \Wt(x)=\frac{(x-\delta)^2U(x)-2(x-\delta)\gamma V(x)-\gamma^2 W(x)}
{x-\l_f},
\end{equation*}
because the numerator of the above right hand side is monic of degree
$g+2$ and vanishes at the roots of $W$ as well as at $x=\l_f$. By
interpolation at the zeros of $\Wt$ we find that
\begin{equation*}
  \Vt(x)=\frac{(x-\delta)(\delta-\l_f)U(x)+(2\delta-\l_f-x)\gamma
          V(x)+\gamma^2 W(x)}{\gamma(x-\l_f)}.
\end{equation*}
We recover our formulas (\ref{W}) and (\ref{V}) (hence also (\ref{U}))
by taking $\gamma=\a$ and $\delta=\lambda-\a\b$).
%
  \subsection{Generalized Jacobians (even case)} %
%
\label{gen_ev_par}
In this case phase space $\Mh_g$ is given by the space of triples of polynomials
$(U(x),V(x),W(x))$ with the following degree constraints
\begin{align*}
  U(x)&=x^{g+1}+U_0x^{g}+\ldots+U_g, \\
  V(x)&=V_0x^{g}+\ldots+V_g, \\
  W(x)&=x^{g+1}+W_0x^{g}+\ldots+W_g.
\end{align*}
In this case the spectral curve is of the form $y^2=f(x)$ where
$f(x)=U(x)W(x)+V^2(x)$ is monic of degree $2g+2$. When $f$ is irreducible the
corresponding fiber of the moment map $\hat\chi$ (which is given as in the other
cases by $\hat\chi(L(x))= -\det L(x)$) is an affine part of $\Sym^{g+1}\Cb_f$; this
is shown by associating to $(U(x),V(x),W(x))\in\chi^{-1}(f)$ the divisor
$D=\sum_{i=1}^{g+1}(x_i,y_i)$, where $x_i$ are the roots of $U(x)$ and
$y_i=V(x_i)$. We choose a section $\xi$ and we let $\xi(f)=(\l_f,\mu_f,f)$. For a
generic point $\UVW\in\Mh_g$ we consider the function
\begin{equation*}
  F(x,y)=\frac{(y+V(x))(x-\a_1)\pm U(x)(x-\a_2)}{U(x)(x-\l_f)},
\end{equation*}
where $\a_1$ and $\a_2$ satisfy the following linear equation (zero of the
numerator of $F$ at the point $(\l_f,-\mu_f)$):
\begin{equation*}
  (-\mu_f+V(\l_f))(\l_f-\a_1)\pm U(\l_f)(\l_f-\a_2)=0.
\end{equation*}
$F(x,y)$ has $\D+(\l_f,\mu_f)=\sum_{i=1}^{g+1}(x_i,y_i)+(\l_f,\mu_f)$ as its polar
divisor and vanishes at infinity. It follows that the other zeros of $F(x,y)$ give a
divisor $\Dt$ which is linearly equivalent to the divisor $\D$ which is associated to
$\UVW$, up to a shift over $(\l_f,\mu_f)-\infty_f$. It leads to the following
formulas for the map $B_\xi$
\begin{align*}
  \Ut(x)&=\frac{U(x)(x-\a_2)^2\pm 2V(x)(x-\a_1)(x-\a_2)-W(x)(x-\a_1)^2}
{C(x-\l_f)}\,,\\
  \Vt(x)&=\frac1{C(x-\l_f)}\left[\pm (x-\a_2)(x-\a_4)U(x)\right.\\
        &\left.+((x-\a_2)(x-\a_3)+(x-\a_1)(x-\a_4))V(x)\mp (x-\a_1)(x-\a_3)
W(x)\right]\,,\\
  \Wt(x)&=\frac{-U(x)(x-\a_4)^2\mp 2V(x)(x-\a_3)(x-\a_4)+W(x)(x-\a_3)^2}
{C(x-\l_f)}\,,
\end{align*}
where
\begin{equation}
  C=2(\a_1-\a_2)+U_0\pm 2V_0-W_0
\end{equation}
and
\begin{equation}
  \a_3=\a_1-C\,\frac{\a_1-\l_f}{\a_1-\a_2}\,, \qquad
  \a_4=\a_2-C\,\frac{\a_2-\l_f}{\a_1-\a_2}\,.
\end{equation}
The above transformation can be rewritten in the form of the matrix equation
$M(x)L(x)=\Lt(x)M(x)$ with the following matrix $M$:
\begin{equation}
  M(x)=\mat{x-\a_3}{\pm (x-\a_4)}{\pm(x-\a_1)}{x-\a_2},
\end{equation}
{}where the variables $\a_1,\ldots,\a_4$ are given by
\begin{equation}
  \a_i=\l_f+\frac{(\epsilon_i C-U_0\mp 2V_0+W_0)((-1)^{i-1}C
-\Ut_0\pm 2\Vt_0+\Wt_0)}
                 {4C},
\end{equation}
with $\epsilon_i=1$ for $i=1,2$ and $\epsilon_i=-1$ otherwise.

Let us now turn to poissonicity and spectrality. For every polynomial
$\varphi$ of degree at most $g+1$ we find a Poisson structure $\Pb^\varphi$
which is given formally by precisely the same formulas as in the case
considered in Paragraph \ref{gen_par}. We can see from the above formulas
that it will be much easier to do further calculations if we make a simple
similarity transform:
\begin{equation}
  M(x)\mapsto SM(x)S^{-1},\qquad L(x)\mapsto SL(x)S^{-1},
\end{equation}
where
\begin{equation}
  S=\mat{1}{\pm 1}{1}{\mp 1}.
\end{equation}
Let us denote the transformed matrices $L(x)$ and $M(x)$ by small letters
$\ell(x)$
and $m(x)$, respectively:
$$
\ell(x)=SL(x)S^{-1},\qquad \tilde \ell(x)=S\Lt(x)S^{-1},
 \qquad m(x)=SM(x)S^{-1},
$$
and correspondingly,
$$
\ell(x)=\mat{v(x)}{w(x)}{u(x)}{-v(x)},\qquad
\tilde \ell(x)=\mat{\vt(x)}{\wt(x)}{\ut(x)}{-\vt(x)}.
$$
The triple of new polynomials is as follows:
\begin{align*}
  u(x)&=u_0x^{g}+\ldots+u_g, \\
  v(x)&=\pm x^{g+1}+v_0x^{g}+\ldots+v_g, \\
  w(x)&=w_0x^{g}+\ldots+w_g,
\end{align*}
and the matrix $m(x)$ has the following form:
$$
m(x)=\frac1{2C} \mat{4(C(x-\l)+w_0\ut_0)}{\pm 2Cw_0}
{\pm 2 C\ut_0}{C^2}.
$$
Note that the determinant of the matrix $m(x)$, as well as of
the matrix $M(x)$, is expressed in terms of $C$:
$$
\det M(x) = \det m(x)=C(x-\l).
$$
Suppose now that  the polynomial $\varphi(x)$ is monic and has distinct roots
$a_0,\dots,a_g$ and consider the map defined by
\begin{equation*}
  \frac{1}{\varphi(x)}(u(x),v(x),w(x))=\left(\sip\frac{f_i}{x-a_i},
  \pm1+\sip\frac{h_i}{x-a_i},\sip\frac{e_i}{x-a_i}\right).
\end{equation*}
It is an isomorphism between $\Mh_g$, equipped with the Poisson structure
$\Pb^\varphi$, and the direct sum of $g+1$ copies of the Lie-Poisson algebra of
$\sllie(2)$. Notice that $u_0=\sip f_i$ and $w_0=\sip e_i,$ so that $m(x)$ depends
only on variables $e_i$ and $\tilde f_i$. Therefore we take $(e_i,\tilde f_i)$,
$i=0,\ldots,g$, as independent variables. Then, it is easy to find the following
formulas for the variables $(h_j,\tilde h_j)$, $j=0,\ldots,g$:
\begin{align}
  C^2h_j^2\mp 4C\ut_0 e_j h_j + 4 e_j(C(a_j-\l)\tilde f_j+\ut_0^2e_j)-C^2C_j=0,
\notag\\
  C^2\tilde h_j^2\mp 4Cw_0 \tilde f_j \tilde h_j + 4 \tilde f_j(C(a_j-\l)e_j+
       w_0^2\tilde f_j)-C^2C_j=0,\label{nu1}\\
  C(h_j-\tilde h_j)=\pm 2 (\ut_0 e_j-w_0\tilde f_j).\notag
\end{align}
As for the compatibility conditions:
$$
e_k \frac{\partial h_j}{\partial e_k}=
e_j \frac{\partial h_k}{\partial e_j}\,,\quad
\tilde f_k \frac{\partial \tilde h_j}{\partial \tilde f_k}=
\tilde f_j \frac{\partial \tilde h_k}{\partial \tilde f_j}\,,
\quad
\tilde f_k \frac{\partial h_j}{\partial \tilde f_k}=
e_j \frac{\partial \tilde h_k}{\partial e_j}\,.
$$
we have from (\ref{nu1}) that
$$
\frac{\partial h_j}{\partial e_k}=\frac{\partial \tilde h_j}{\partial
\tilde f_k}=0,\qquad j\neq k.
$$
which leads at once to the first two equations and to the third equation for
$j\neq k$. The proof of the third equation for $i=k$ is easy by direct
computation.
The spectrality property also holds, as one easily verifies the following
 formulas:
$$
e_j \frac{\partial \hat \mu}{\partial e_j}=
-\frac{\partial h_j}{\partial \l}\,,\qquad
\tilde f_j \frac{\partial \hat \mu}{\partial \tilde f_j}=
-\frac{\partial \tilde h_j}{\partial \l}\,,
$$
where $\hat\mu={\mu}/{\varphi(\l)}.$

We finish by computing the continuum flows, obtained by taking the limit
$t\to 0$ of
the family of sections $\xi_t$ given by $\l=1/t$ and
$\mu=\mp(1+(U_0+W_0)t/2+O(t^2))/t^{g+1}$. In order for the limit to exist
we must
take the Casimir $C$ of the form $C'-4\l$, where $C'$ does not depend on
$\l$. Then
\begin{align*}
  a_1&=\frac{C'-2U_0+2(\pm1-1)V_0+2W_0}4,\\
  a_2&=\frac2t-\frac{C'-2(1\pm1)V_0}4,
\end{align*}
and in the limit our \BT s lead, as in the other cases, to a vector field which has
the Lax form $L'(x)=[L(x),N(x)]$, where $N(x)$ is given (up to a constant
factor $1/8$) by
\begin{equation*}
  \mat{(2\pm2)V_0}{\pm(4x-C'-2U_0-(2\mp2)V_0+2W_0)}
{\pm(4x-C'+2U_0+(2\mp2)V_0-2W_0)}
  {-(2\pm2)V_0}.
\end{equation*}
In terms of $l(x)$ this becomes $l'(x)=[l(x),n(x)]$, where
$n(x)=VN(x)V^{-1}$ is given by
\begin{equation*}
  n(x)=\pm\frac18\mat{4x-C'}{4w_0}{4u_0}{C'-4x}.
\end{equation*}
The above vector fields is the analog of the vector field $X_\infty$ of the
Mumford system. The analogs of the vector fields $X_a,\,a\in\P^1$ are
constructed in the same way as in the other cases.

%
  \subsection{Geodesic flow on SO(4)} %
%
We now look at the case of an integrable geodesic flow on {\bf SO(4)},
whose underlying metric appears as metric II in the classification of
integrable geodesic flows on {\bf SO(4)}. In suitable coordinates, the
basic vector field $X_1$ of this \aci\ is given by the
differential equations
%
%
\begin{equation*}
  \begin{array}{lll}
     \dot z_1  = 2 z_5z_6,\qquad &\dot z_2 = 2z_3z_4,       \qquad
 &\dot z_3 =
z_5(z_1+z_4),\\
     \dot z_4  = 2z_2z_3, \qquad  &\dot z_5 = z_3(z_1+z_4), \qquad
 &\dot z_6 = 2z_1z_5.\\
  \end{array}
\end{equation*}
and it admits the following quadratic first integrals:
\begin{align}
   H_1 & = z_3^2-z_5^2,\notag\\
   H_2 & = z_1^2-z_6^2,\notag\\
   H_3 & = z_2^2-z_4^2,\label{H_i}\\
   H_4 & = (z_1+z_4)^2+4(z_3^2-z_2z_5-z_3z_6).\notag
\end{align}
Following \cite{BV} we let
\begin{equation*}
u(x)=x^2+\left(\frac{z_1+z_2+z_4+z_6}{2(z_3-z_5)}-1\right)x-
\frac{z_2+z_4}{2(z_3-z_5)},
\end{equation*}
and we let $v(x)$ be the polynomial of degree at most 1, characterized
by
\begin{equation*}
    v(0)=u(0)(z_1+z_4+2z_3),\quad v(1)=u(1)(z_1+z_4+2z_5).
\end{equation*}
This map associates to any point $P$ in $\C^6$ an unordered pair of points
on the algebraic curve
\begin{equation}\label{curve}
  \Gamma:y^2=f(x)=x(1-x)\left[4x^3h_1-(4h_1+h_4)x^2+(h_4-h_3-h_2)x
+h_3\right],
\end{equation}
where $h_i$ denotes the value of $H_i$ at $P$.  Notice that the polynomial
$f$ which
defines $\Gamma$ is not monic, its leading term being dependent on the integrals. As
a consequence, the polynomial $w$, defined by $ w(x)={f(x)-v^2(x)}/{u(x)}$,
will not
be monic and the map does \emph{not} define a map to the Mumford system (indeed, for
most of the Poisson structures of this system this leading term is not even a
Casimir). For future use, notice that $w(0)=-u(0)(z_1+z_4+2z_3)^2$ and
$w(1)=-u(1)(z_1+z_4+2z_5)^2$, because $f$ has $0$ and $1$ as roots.
 Conversely,
given three such polynomials $u,v,w$ which satisfy $v^2(x)+u(x)w(x)=f(x)$,
where $f$
has the above form (\ref{curve}), the corresponding point
$(z_1,\dots,z_6)\in\C^6$ is
reconstructed by using the following formulas.
\begin{align}
  z_3-z_5&={1\over2}\left({v(0)\over u(0)}-{v(1)\over u(1)}\right),\notag\\
  z_2+z_4&=\left({v(1)\over u(1)}-{v(0)\over u(0)}\right)u(0),\label{trans}\\
  z_1+z_6&=\left({v(0)\over u(0)}-{v(1)\over u(1)}\right)u(1),\notag
\end{align}
in addition to the first three equations in (\ref{H_i}).

In order to construct \BT s for this system we consider, for a fixed point
$P\in\C^6$, the following rational function
\begin{equation}\label{F_SO}
  F(x,y)=\frac{y+v(x)+\b u(x)}{u(x)(x-\l_f)},
\end{equation}
and we demand that the numerator of $F$ vanishes at the point $(\l_f,-\mu_f)$, as in
the case of the Mumford system. It leads to
\begin{align}
  \ut(x)&=\frac{\b^2u(x)+2\b v(x)-w(x)}{-4h_1(\l_f-x)},\notag\\
  \vt(x)&=\frac{(\b^3-4h_1\b(x-\l_f))u(x)+(2\b^2-4h_1(x-\l_f))v(x)-\b w(x)}
               {4h_1(x-\l_f)},\label{SOBT}
\end{align}
the value of $\wt(x)$ is not needed for the computation. Writing (\ref{trans}) in
terms of tilded variables and substituting (\ref{SOBT}) in it we find
\begin{align*}
  \frac{\zt_3-\zt_5}{z_3-z_5}&=2(z_3+z_5)\left(\frac{\l_f}{z_1+z_4+2z_3+\b}-
  \frac{\l_f-1}{z_1+z_4+2z_5+\b}\right),\\
  \frac{\zt_2+\zt_4}{z_2+z_4}&=-\frac14\frac{\zt_3-\zt_5}{z_3-z_5}
  \frac{(z_1+z_4+2z_3+\b)^2}{h_1\l_f},\\
  \frac{\zt_1+\zt_6}{z_1+z_6}&=-\frac14\frac{\zt_3-\zt_5}{z_3-z_5}
  \frac{(z_1+z_4+2z_5+\b)^2}{h_1(\l_f-1)}.\\
\end{align*}
Since the map preserves the Hamiltonians the above three expressions
are (in that order) also equal to
\begin{equation*}
  \frac{z_3+z_5}{\zt_3+\zt_5},\qquad
  \frac{z_2-z_4}{\zt_2-\zt_4},\qquad
  \frac{z_1-z_6}{\zt_1-\zt_6},
\end{equation*}
so that the above equations can be solved linearly in terms of the variables
$\zt_i$. The Poisson matrix of a Poisson structure for this system is given by
\begin{equation*}
  \left(
    \begin{array}{cccccc}
      0&z_6&-z_5&0&-z_3&z_2-2z_5\\
       -z_6&0&0&z_6-2z_3&0&-z_1-z_4\\
       z_5&0&0&-z_5&0&0\\
       0&2z_3-z_6&z_5&0&z_3&-z_2\\
       z_3&0&0&-z_3&0&0\\
       2z_5-z_2&z_1+z_4&0&z_2&0&0
    \end{array}
  \right).
\end{equation*}

If $\l$ depends on the Casimirs of this Poisson structure only, then the above map is
a Poisson map, so it is a \BT; moreover it has the spectrality property. This can be
verified directly by computing the brackets $\{\tilde z_i, \tilde z_j\}$ and
verifying the compatibility relations. Alternatively one uses the fact that the map
which is induced on the triples of polynomials $\uvw$, as above, is a \BT\ for an
\aci\ obtained by removing in the Mumford system the restriction that the polynomial
$w$ be monic (the Poisson structures are obtained from (\ref{PB_len}) by replacing
$\sigma\t\sigma$ with $\bar w-\sigma\t\sigma$, where $\bar w$ denotes the leading
coefficient of $w(x)$). It suffices then to verify that the map which sends
$(z_1,\dots,z_6)$ to $\uvw$ is a Poisson map and has the spectrality property when
one takes on the target space the Poisson structure corresponding to
$\varphi(x)=x(x-1)$.

  \subsection{The H\'enon-Heiles potential} 

\label{HH_par}
In this paragraph we show on an example how one gets \BT s for \aci s whose generic
level set of the integrals is a finite cover of a Jacobian. We do this by lifting the
\BT\ for the underlying family of Jacobians to the cover; since such a lifting is not
unique we get, in general, an implicitly defined correspondence, rather than an
explicit map.

We treat the case of the H\'enon-Heiles system, which is given by the
following Hamiltonian on $\C^4$, equipped with the standard
symplectic structure,
\begin{equation*}
  H=\frac12\left(p_1^2+p_2^2\right)+8q_2^3+4q_1^2q_2.
\end{equation*}
A first integral is given by
\begin{equation*}
  F=-q_2p_1^2+q_1p_1p_2+q_1^2(q_1^2+4q_2^2).
\end{equation*}
We use the map defined by
\begin{align}
  u(x)&=x^2-2q_2x-q_1^2,\notag\\
  v(x)&=\frac i{\sqrt{2}}(p_2x+q_1p_1),\label{uvwHH}\\
  w(x)&=x^3+2q_2x^2+(q_1^2+4q_2^2)x-\frac{p_1^2}2,\notag
\end{align}
which is a morphism to the Mumford system, the latter being equipped
with the Poisson
structure corresponding to $\varphi(x)=x$. It follows from the results of
 Section
\ref{Mumford} that for any constant $\l\in\C$ we get a \BT, given by
$\Lt=MLM^{-1}$,
where
\begin{equation*}
  L(x)=\mat{\frac i{\sqrt{2}}(p_2x+q_1p_1)}
            {x^3+2q_2x^2+(q_1^2+4q_2^2)x-\frac{p_1^2}2}
            {x^2-2q_2x-q_1^2}
            {-\frac i{\sqrt{2}}(p_2x+q_1p_1)}
\end{equation*}
and
\begin{equation*}
  M(x)= \mat{\b }{x-\l_f+\b^2}{1}{\b },
  \quad\hbox{where}\quad
  \b=\frac{\sqrt2\mu_f-i(p_2\l_f+q_1p_1)}{\sqrt2(\l^2-2q_2\l-q_1^2)}.
\end{equation*}
Also $\mu_f^2=f(\l_f)$ with
\begin{equation*}
  f(x)=u(x)w(x)+v^2(x)=x(x^4-hx-g),
\end{equation*}
where $h$ and $g$ are the values of $H$ and $G$ at the point
$(q_1,q_2,p_1,p_2)$. Poissonicity and spectrality are a consequence
of the fact that
the map $ (q_1,q_2,p_1,p_2)\to(u,v,w)$, given by (\ref{uvwHH}) is
a Poisson map. One
notices that in this case one does not get explicit formulas for
 $\tilde q_1,\tilde
q_2,\tilde p_1,\tilde p_2$ but for $\tilde q_1^2,\tilde q_2,
\tilde q_1p_1,\tilde
p_2$, which stems from the fact that the generic level manifolds of the
integrals are
$2:1$ unramified covers of Jacobians.

\section{Concluding remarks}
We have constructed B\"{a}cklund transformations for a large class of integrable
systems. Basically, we have considered four large families of integrable systems that
are of interest in mathematical physics.  Indeed, if we choose the following
parametrization of the generators $(h_j,e_j,f_j)$ of a direct sum
of $g$ or $g+1$
copies of the Lie-Poisson algebra of $\sllie(2)$, in terms of the canonical Darboux
variables (coordinates and momenta),
$(p_j,q_j)$, $\{p_j,q_k\}=\delta_{jk}$:
\begin{equation*}
  h_j=\frac12\,p_jq_j,\qquad
  f_j=\frac12\,q_j^2,\qquad
  e_j=-\frac12\,p_j^2+\frac{2C_j}{q_j^2},
\end{equation*}
then we deal with the following Hamiltonian systems.

\noindent (1) In the case of the Mumford system the Hamiltonian is of the form:
$$
  H=\frac12 \sum_{i=1}^gp_i^2-\sum_{i=1}^g \frac{2C_i}{q_i^2}
    -\frac12 \sum_{i=1}^g q_i^2(a_i+C_0)+\frac14
\left(\sum_{k=1}^g q_k^2\right)^2,
$$
so this case is a generalization of the $g$-dimensional Garnier system.

\noindent (2) For the even Mumford system the Hamiltonian function describes the
motion of a particle in a potential of order 6:
\begin{align*}
  H&=\frac12 \sum_{i=1}^gp_i^2-\sum_{i=1}^g \frac{2C_i}{q_i^2}
     -\frac12 \sum_{i=1}^g (a_i^2+a_iC_{-1}+C_0)q_i^2\\
   &\qquad  +\frac14
     \left(\sum_{k=1}^gq_k^2\right)\sum_{i=1}^g(C_{-1}+2a_i)q_i^2-\frac18
     \left(\sum_{k=1}^gq_k^2\right)^3.
\end{align*}
\noindent (3) In the odd generalized case we have an integrable system
with linear potential
$$
 H=\frac12 \sum_{i=0}^gp_i^2-\sum_{i=1}^g \frac{2C_i}{q_i^2}
   +\frac12 q_0,
$$
\noindent
(4) In the even generalized case we have a $g$-dimensional
harmonic oscillator
$$
 H=\frac12 \sum_{i=0}^gp_i^2-\sum_{i=0}^g \frac{2C_i}{q_i^2}
    -\frac12 \sum_{i=0}^g q_i^2.
$$

In other words we have showed how to construct in a systematic way B\"{a}cklund
transformations for integrable systems linearisable on hyperelliptic Jacobians or
generalized hyperelliptic Jacobians. Since for many classical integrable systems it
is known how to embed them into Mumford systems \cite{Van1}, our construction
produces many new integrable discretizations of Liouville integrable systems, such as
the Kowalevski, Goryachev-Chaplygin and Euler tops, Toda lattices and the Gaudin magnet.

\bibliographystyle{amsplain}
\providecommand{\bysame}{\leavevmode\hbox to3em{\hrulefill}\thinspace}

\end{document}